# Natural-mixing guided design of refractory high-entropy alloys with as-cast tensile ductility

Shaolou Wei[1], Sang Jun Kim[2], Ji Yun Kang[1], Yong Zhang[3], Yongjie Zhang[4],

Tadashi Furuhara[4], Eun Soo Park[2], and Cemal Cem Tasan[1,*]

[1.] *Department of Materials Science and Engineering, Massachusetts Institute of Technology, Cambridge, MA 02139, USA*
[2.] *Department of Materials Science and Engineering, Seoul National University, Seoul 08826, Republic of Korea*
[3.] *Materials Research Laboratory, Massachusetts Institute of Technology, Cambridge, MA 02139, USA*
[4.] *Institute for Materials Research, Tohoku University, Sendai 980-8577, Japan*

*Corresponding author: Cemal Cem Tasan (tasan@mit.edu)

**Multi-principal-element metallic alloys**[1,2] **have created a growing interest that is unprecedented in metallurgical history, in exploring the property limits of metals**[3–7] **and the governing physical mechanisms**[8–11]**. Refractory high-entropy alloys (RHEAs) have drawn particular attention due to their (i) high melting points and excellent softening-resistance, which are the two key requirements for high-temperature applications**[12]**; and (ii) compositional space, which is immense even after considering cost and recyclability restrictions**[13]**. However, RHEAs also exhibit intrinsic brittleness and oxidation-susceptibility, which remain as significant challenges for their processing and application. Here, utilizing natural-mixing characteristics amongst refractory elements, we designed a $Ti_{38}V_{15}Nb_{23}Hf_{24}$ RHEA that exhibits >20% tensile ductility already at the as-cast state, and physicochemical stability at high-temperatures. Exploring the underlying deformation mechanisms across multiple length-scales, we observe that a rare $\beta'$ precipitation strengthening mechanism governs its intriguing mechanical response. These results also reveal the effectiveness of natural-mixing tendencies in expediting HEA discovery.**

The attention that RHEAs received increased exponentially since the reporting of the first examples with single-phase body-centered cubic (BCC) structures [14], owing to the yield strength preservation tendency they exhibit at elevated temperatures. Since then, numerous alloys have been assessed both theoretically and experimentally, yet, the efficiency of this compositional search has been low. Apart from the immense compositional space that drastically hinders accurate phase diagram computations, there are several other serious obstacles[14,15]: (i) governed by the milder discrepancy between ideal shear strength and ideal tensile strength for cleavage formation[16], the majority of the reported RHEAs demonstrate negligible tensile ductility and are intrinsically brittle at ambient temperature; (ii) extensive homogenization treatments are often inevitable due to the sluggish diffusion kinetics of refractory elements[17]; and (iii) the presence of catastrophic oxidation at intermediate temperatures (800-1000 ºC) retards the application of classical hot-processing treatments[18]. Due to these challenges and the absence of sufficiently-sophisticated thermodynamic-kinetic fundamentals, the *RHEA rush* has led to few alloys that exhibit application-worthy mechanical performances[19,20]. We reveal in this work that by following an approach that exploits the natural mixing characteristics amongst refractory elements to minimize casting segregation, it is possible to expediently guide the RHEA composition search, and thereby to achieve excellent strength-ductility-high temperature stability combinations.

The first step of our composition searching strategy involves quantitative elemental partition assessment at the meso-scale, by probing the largest single-phase region inherited from natural mixing amongst refractory elements. To achieve this, a nine-component master RHEA consisting of equal atomic portions of Ti, V, Cr, Zr, Nb, Mo, Hf, Ta, and W was cast *via* arc-melting. As displayed in Fig. 1 a, four predominant phase-separated zones develop in its microstructure, enriched respectively in Ti, Mo, Cr, and Hf (see the markers in **Fig. 1 a**). With the aid of energy dispersive spectroscopy (EDS) elemental mapping and point analyses, it is realized that the largest single-phase region that spans almost over the full microstructure demonstrates a composition of $Ti_{38}V_{15}Nb_{23}Hf_{24}$ at. % (i.e. considering alloying elements present >10 at. % as principal constituents). Next, four specimens with this exact naturally-guided compositions excerpted from the master RHEA were produced again by arc-melting, and the corresponding as-cast microstructures are comparatively revealed in **Fig. 1 b**. Surprisingly, in contrast to the gigantic dendrites or lamellae present in the three other compositions, the $Ti_{38}V_{15}Nb_{23}Hf_{24}$ combination shows a near-equiaxed grain morphology with negligible dendritic microstructure, implying thermodynamically preferred mixing characteristics and a relatively minor amount of solute segregation. After recrystallization processing (details in "**Methods**"), this $Ti_{38}V_{15}Nb_{23}Hf_{24}$ RHEA displays a fully-equiaxed grain morphology with a single BCC-phase constitution ($a = 3.318$ Å) at the resolution limit of electron back scatter diffraction (EBSD) and synchrotron X-ray techniques (**Fig. 1 c**). EDS elemental mapping across multiple grain boundaries also evidences the spatially uniform distribution of the four alloying elements at the meso-scale.

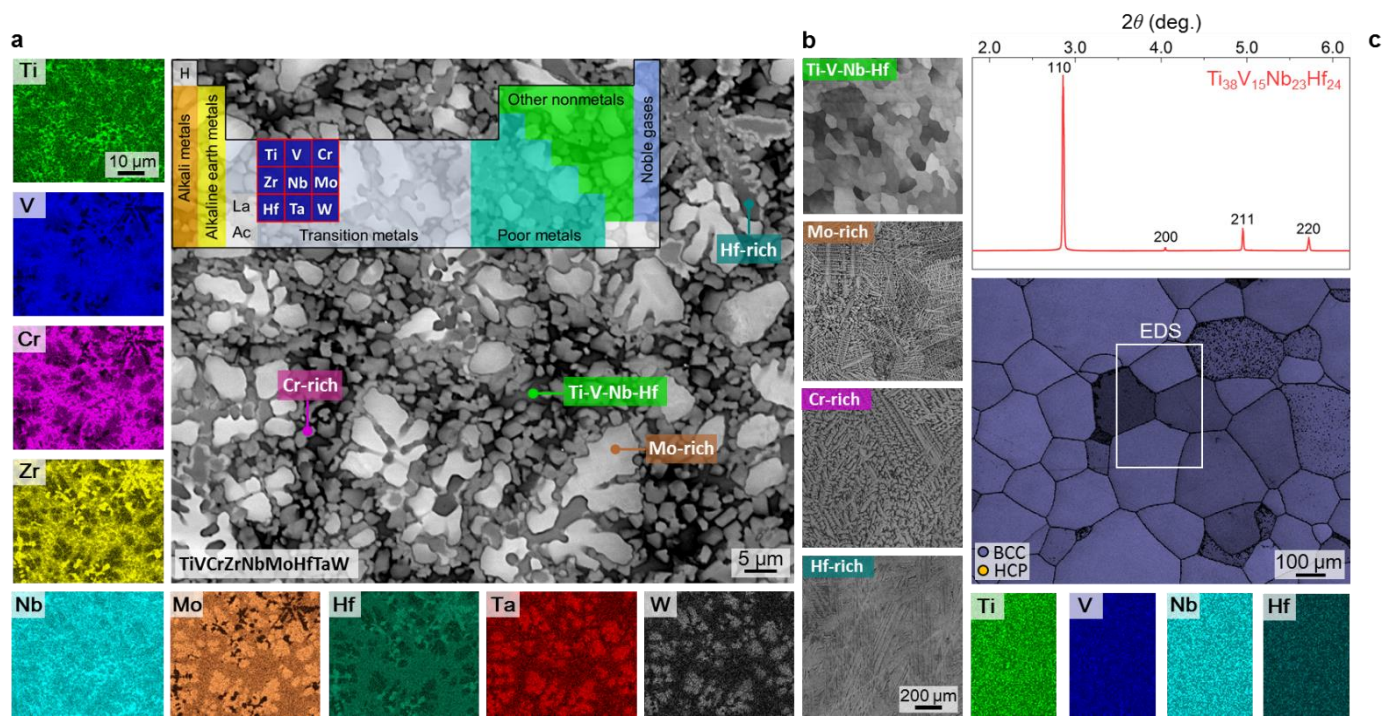

**Fig. 1| The strategy of composition search amongst refractory elements. a,** as-cast microstructure of a master RHEA consisting of the nine refractory elements in equal atomic portion. The corresponding EDS mapping results reveal the distinctive presence of four phase-separated regions, respectively enriched in Ti, Mo, Cr, and Hf. **b,** as-cast microstructures of the four alloys designed using the compositions measured from the corresponding regions in **a** (more details are provided in supplementary **Fig. S1** and **Tab. S1**). **c,** microstructural characterization of the recrystallized $Ti_{38}V_{15}Nb_{23}Hf_{24}$ RHEA whose composition was inherited from the largest single-phase region in **a** (denoted as "Ti-V-Nb-Hf"). Sub-figures presented from top to bottom in **c** are synchrotron X-ray diffraction patterns, EBSD phase map, and EDS elemental distribution assessment of a selected region. All the meso-scale characterizations indicate that the $Ti_{38}V_{15}Nb_{23}Hf_{24}$ RHEA preserves a single-phase BCC structure with a lattice parameter $a = 3.318$ Å.

We next examine the ambient temperature tensile properties of the $Ti_{38}V_{15}Nb_{23}Hf_{24}$ RHEA, starting with the as-cast state. As seen in **Fig. 2 a**, the mechanical response in the as-cast state, even without any further optimization, is strikingly well. A yield strength of ~800 MPa is achieved while preserving a desirable fracture elongation of ~21.6 %. In fact, further homogenization and recrystallization treatments bring about only subtle variations in yield strength (~802 MPa), ultimate tensile strength (regarded the same as the yield strength), and fracture elongation (~22.5 %). By comparing the tensile properties of this alloy and its equiatomic counterpart (discussed in more detail below), with other RHEAs proposed in literature, it is observed that these new alloys exhibit excellent specific yield strength-tensile ductility combinations, even in these un-optimized states (inset of Fig. **2 a** and supplementary **Fig. S3**).

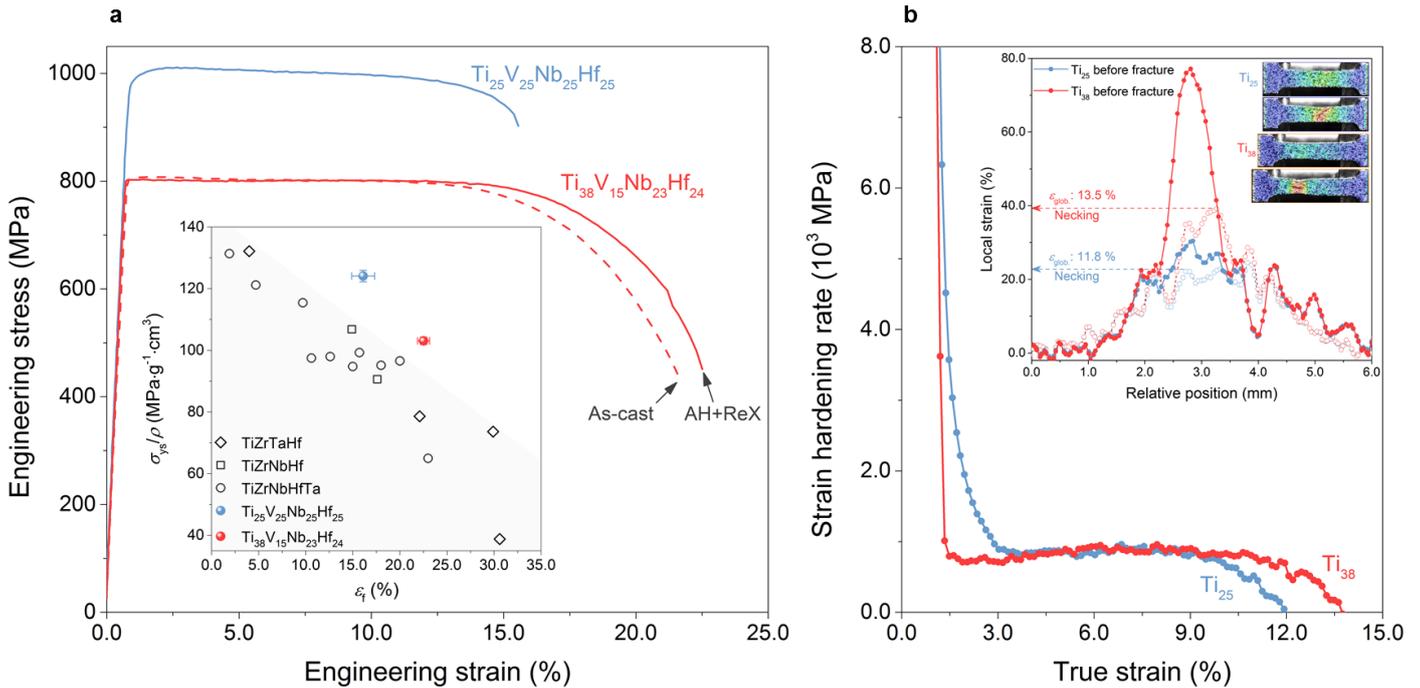

**Fig. 2| Mechanical behavior of the designed $Ti_{38}V_{15}Nb_{23}Hf_{24}$ RHEA and its equiatomic variant (detailed characterizations are revealed in supplementary Fig. S2). a,** uniaxial tensile properties at ambient temperature. The strength-ductility combination of $Ti_{38}V_{15}Nb_{23}Hf_{24}$ RHEA is shown both in as-cast state and after homogenization-recrystallization treatment (marked as "AH-ReX"). An improved yield strength along with a milder yielding-inflection response is observed in the equiatomic variant which is mainly due to the increase in the size and fraction of the nano-precipitates (supplementary **Figs. S11 and S12**). Inset of **a** displays an Ashby chart showing the unique advantages of these two RHEAs exist in their desirable specific yield strength ($\sigma_{ys}/\rho$)-fracture elongation ($\varepsilon_f$) synergy compared with the literature (three other complementary comparisons and referencing details are provided in supplementary **Fig. S3 and Tab. S2**). **b,** strain hardening rate plots with local strain profiles. Since the actual onset of necking in metallic alloys that exhibit a plateau-like plastic response may often contradict with the prediction of Considère criterion ($d\sigma/d\varepsilon > \sigma$) [21], we therefore determine the necking incipience point by performing digital image correlation measurement (details see "**Methods**"). Strain maps presented in the inset are the two-dimensional local strain distribution of the two RHEAs at the onsets of necking and one frame right before fracture takes place.

Note that the main goal of the present work is to report the feasibility of the design strategy based on natural mixing tendencies of refractory alloying elements. The properties that are achieved in an un-optimized, as-cast state, already demonstrate the success of this approach. That stated, further property improvements can be realized by microstructural optimization, and/or by fine-tuning the alloy composition *via* systematically exploring the TiVNbHf alloy system. As an example, we showcase the tensile properties of an equiatomic ramification (namely, the $Ti_{25}V_{25}Nb_{25}Hf_{25}$ at. % RHEA, supplementary **Fig. S2**) which exhibits a yield strength of ~1004 MPa with ~16.1 % fracture elongation. Clearly, a large property space can be accessed by exploring different microstructure and composition variations of TiVNbHf.

In addition to the promising property combinations, these two RHEAs also exhibit intriguing deformation characteristics, suggesting the involvement of new deformation micro-mechanisms. The starting point of our investigation of the underlying micro-mechanisms is the comparison of the strain hardening response of these two RHEAs. A closer examination of **Fig. 2 b** indicates that although these two RHEAs exhibit a similar, moderate strain hardening rate during stable plastic flow (plateau-like engineering stress-strain curves in **Fig. 2 a**), notable differences exist in (i) initial yielding response (true strain<3.0 % portion in **Fig. 2b**); and (ii) post-necking elongation (inset of **Fig. 2 b** and fractography in supplementary **Fig. S4**). To explain the latter, we have carried out a series of strain rate jump tests ($1 \times 10^{-4} \sim 1 \times 10^{-3}$ s$^{-1}$). The results shown in supplementary **Fig. S5** reveal that, compared to its equiatomic counterpart, the $Ti_{38}V_{15}Nb_{23}Hf_{24}$ RHEA exhibits a higher strain rate sensitivity – the dominant factor that controls post-necking elongation, based on the Hutchinson-Neale non-linear analysis[22]. The explanation of the former case, i.e. the difference in the yielding response, requires more dedicated analyses. We propose that the sharp yielding point together with the moderate strain hardening rate seen in the $Ti_{38}V_{15}Nb_{23}Hf_{24}$ RHEA results from a unique dislocation channeling mechanism, which is discussed in detail next.

As shown in **Fig. 3 a** for the local strain level of ~10.0 %, rectilinear surface steps can be observed with the initiation of plasticity on a deformed $Ti_{38}V_{15}Nb_{23}Hf_{24}$ RHEA sample, the surface of which is polished priori to straining. These steps vary in width (0.4-1 μm), and are deeper than classical slip steps that would be present at these strain levels. Such features are also observed using electron channeling contrast imaging (ECCI), at the cross section of fractured samples (polished after straining) (**Fig. 3 b**). ECCI reveals a discernable bright band (termed as dislocation channel in the following) with ~0.65 μm width (**Fig. 3 c1**) developing from grain boundary to its interior at a local strain level of ~3.0 %. Classically, this sort of phenomenon signifies the presence of either stress-assisted martensitic transformation (especially if accompanied with plateau-like engineering stress-strain curves)[23] or mechanical twinning[24]. Surprisingly, the corresponding EBSD inverse pole figure (IPF) and phase maps (**Fig. 3 c2**) clearly prove the absence of martensitic phase formation or characteristic misorientation change (**Fig. 3 c1**) across the band (also validated with EBSD at a higher local strain level and in-situ synchrotron X-ray measurement, see supplementary **Figs. S6 and S7**.). However, higher grain reference orientation deviation (GROD) values present within the band (**Fig. 3 c2**), imply that extensive dislocation-mediated plasticity has taken place within this confined region. In contrast, higher resolution ECCI micrographs (**Fig. 3 b**, above) resolve that dislocations outside the band demonstrate bowed configurations, suggesting the existence of nano-scale heterogeneous sites that pin the dislocations, which as a result, eminently suppresses their mobility (**Fig. 3 b**). To eliminate any potential artifact in ECCI characterization, we have conducted further TEM analyses for validation where such similar features are again observed (**Fig. 3 b**, below).

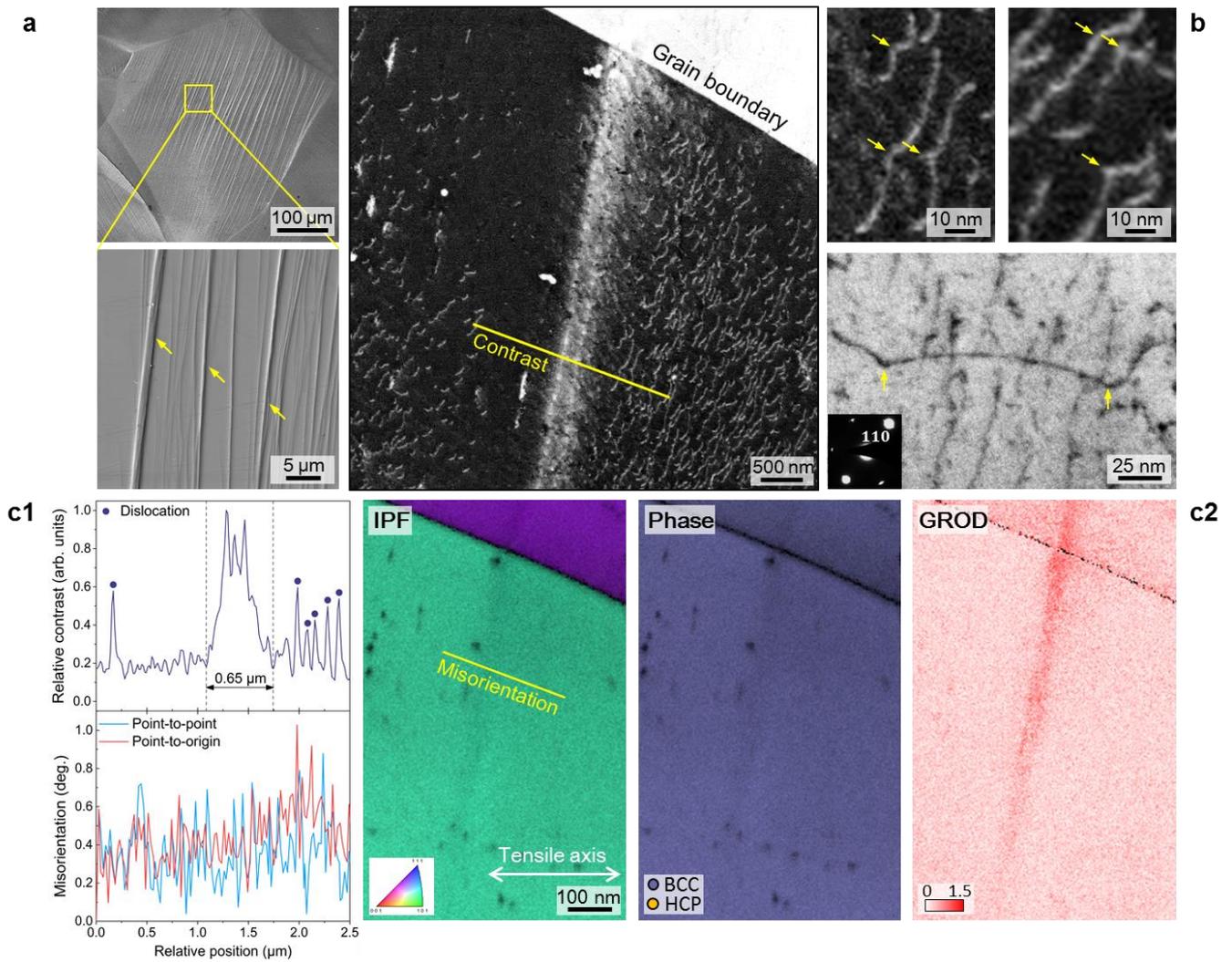

**Fig. 3| Investigation of deformation mechanisms for the Ti$_{38}$V$_{15}$Nb$_{23}$Hf$_{24}$ RHEA at a meso-scale. a,** morphology of the deep rectilinear surface steps at a local strain level of ~10 %. **b,** dislocation configuration characterized by ECCI and TEM. From left to right: low-magnification ECCI micrograph indicating the presence of a discernable channeling band at a local strain level of ~3.0 %; higher-magnification ECCI micrographs showing the bowed dislocation configuration which implies the existence of nano-scale heterogeneities; post-mortem TEM bright field (BF) image validating the ECCI observations; **c1,** quantitative assessment of contrast and local misorientation variation due to the formation of the channeling band. **c2,** EBSD IPF, phase, and GROD maps evidenced although there is no trait of mechanical twinning or phase transformation, significant dislocation-meditated plastic deformation has taken place within the channeling band.

On the basis of these meso-scale observations, the following three hypotheses can be posed regarding the microstructure and the corresponding deformation modules in this alloy: (i) Widely distributed nano-scale heterogeneity exists in the Ti$_{38}$V$_{15}$Nb$_{23}$Hf$_{24}$ RHEA, contributing to the critical stress barrier for plastic incipience. (ii) Plasticity is initiated by the nucleation of dislocation channels, along which these penetrable nano-scale heterogeneities are destructed, leading to an abrupt yielding response. (iii) The following plastic flow with low strain hardening rate arises from the relatively invariant resolved stress necessary for nucleating each channel, of which the strain hardening contribution, however, is still larger than out-of-channel dislocation glide contributions.

To test and validate the mechanistic postulates summarized above, we have carried out multi-probe analyses of the atomic structure of the designed RHEA. High-resolution transmission electron microscope images (HRTEM) together with the corresponding fast Fourier transformed diffraction patterns (**Fig. 4 a**) highlight that in addition to the primary BCC-structured matrix (denoted as β-phase), secondary nano-scale precipitates (denoted as β′-phase) also exist within the $Ti_{38}V_{15}Nb_{23}Hf_{24}$ RHEA. Interestingly, these secondary phases demonstrate a contracted body-centered tetragonal structure with lattice parameters $a = 3.318$ Å and $c = 3.042$ Å. Quantitative atomic probe tomography (APT) assessments (supplementary **Fig. S8**) suggest that these nano-scale precipitates (diameters < 5 nm) are significantly enriched in V, but depleted in Ti (**Fig. 4b** and **c**).

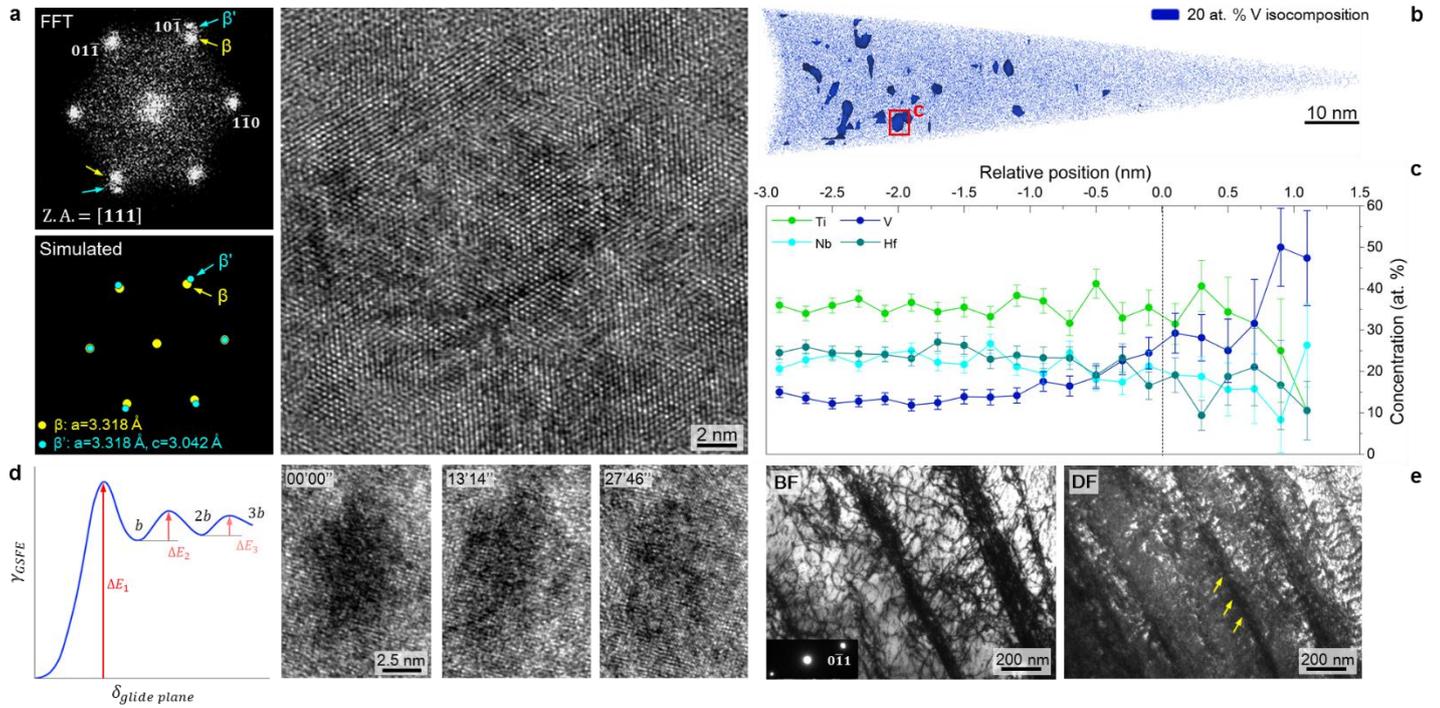

**Fig. 4| Atomic-scale characterization of structural heterogeneity within the $Ti_{38}V_{15}Nb_{23}Hf_{24}$ RHEA. a,** HRTEM micrograph with the corresponding FFT patterns indicate the presence of BCT-structured nanometer scale precipitates. As seen from the measured FFT patterns and the simulated ones (accomplished *via* a Phase Transformation Crystallographic Laboratory software[25]), the two groups of patterns overlap at $1\bar{1}0$ and $\bar{1}10$ spots but splits at the others. Such symmetry features demonstrate that the crystallography of the nano-precipitates yield a BCT structure with a contracted c axis. **b,** APT map excerpted as a 20 at. % V isocomposition surface proves the compositional heterogeneity. **c,** the corresponding poxigram taken along a selected particle in **b** showing the elemental partition characteristics between the BCC-matrix and the BCT-precipitates. **d,** On the left, a qualitative theoretical framework is presented for the nano-precipitate destruction-induced glide plane softening mechanism. Here the generalized stacking fault energy ($\gamma_{GSFE}$)[26] is sketched as a function of displacement ($\delta_{glide\ plane}$) on one specific glide plane. Series of HRTEM micrographs of one such destructed region are shown on the right. With extending electron beam annealing duration, this region undergoes structural recovery that gradually rejuvenates the local lattice periodicity, which supports the proposed precipitate-destruction mechanisms. **e,** TEM BF (close to a two-beam condition) and the corresponding DF micrograph taken from a deformed specimen. Extended information is provided in supplementary **Fig. S13**, proving no mechanical twinning or phase transformation has taken place in such a region.

Considering the geometric and crystallographic characteristics of the nano-precipitates (i.e. their diminutive dimension and the coherency between the precipitates and the matrix) the following mechanism can be proposed to elucidate the origin and energetics of the dislocation channeling effect (**Fig. 4d**, left), and its linkage with the moderate strain hardening rate. The critical interaction here is the one between the precipitate and the first dislocation which manages to penetrate through it (after overcoming an energy barrier of $\Delta E_1$) by shearing a distance of $b$ on its glide plane, destructing the thermodynamically favored local chemical configuration. From an energetic perspective, such an event can be regarded as a procedure that elevates the free-energy of the system, giving rise to glide plane softening[27]. Consequently, the follow-up dislocation(s) that will glide in the same channel will witness an attenuating amount of energy barrier ($\Delta E_1 \rightarrow \Delta E_2 \rightarrow \Delta E_3$ in **Fig. 4d**). This will be the case until the dislocation mean free path decreases-due to multiplication events within the channel-down to a certain threshold below which forest hardening starts to occur. In the aftermath of such a mechanism, one would expect such destructed local sites to exhibit severe lattice periodicity distortion, holding comparatively higher energy states than adjacent microstructural constituents. This enables the design of an *in-situ* experiment to validate the proposed mechanism (**Fig. 4d**, right). The experiment consists of taking a series of HRTEM micrographs of such a site (on a deformed sample) at extended focusing durations, intended to induce beam-annealing effects. The first image shows the investigated region, which is confirmed by diffraction to be a β′ precipitate, the severe lattice periodicity distortion of which is in line with the process proposed above. As seen in the micrograph series shown in **Fig. 4d** (right), this precipitate with deformation-induced local distortion undergoes structural recovery, leading to the observation of improved lattice periodicity. This again supports the proposed mechanism of nano-precipitate destruction. Moreover, lower magnification TEM analyses (**Fig. 4e**) also verify that the highly dislocated regions (confirmed with BF imaging) exhibit a significant contrast diminution (see the corresponding dark field (DF) image), consistent with the proposed mechanism, and the previous literature[28,29].

As mentioned earlier, the exploration of the natural mixing trend amongst refractory elements is highlighting not only one specific RHEA but a whole family of RHEAs (i.e. TiVNbHf alloys). An equiatomic variant presented in Fig. 2 shows the feasibility to peruse more enhanced yield strength without drastically sacrificing ductility *via* fine-tuning the Ti/V ratio. The corresponding change in mechanisms is associated with the decrease of BCC-phase stability which leads to the formation of BCT nano-precipitates with larger size and higher volume fraction (supplementary **Figs. S11 and S12**). Interestingly, from a physicochemical stability perspective, the naturally-predicted $Ti_{38}V_{15}Nb_{23}Hf_{24}$ RHEA does demonstrate significant advantages over its equiatomic ramification. Although the coupled thermal gravity-differential scanning calorimetry (TG-DSC) assessments indicate no eminent signs of phase transformations in both RHEAs up to 1000 °C, a separate 10 h annealing treatment at 1000 °C does prove the formation of hexagonal close packing (HCP)-structured precipitates in the equiatomic variant. The $Ti_{38}V_{15}Nb_{23}Hf_{24}$ RHEA, on the other hand, preserves the major BCC-phase (supplementary **Figs. S14 and S15**). Furthermore, when being subjected to thermal oxidation (600-1000 °C temperature range), no clear trait of catastrophic oxidation is observed for the $Ti_{38}V_{15}Nb_{23}Hf_{24}$ RHEA, while its equiatomic counterpart exhibits severe oxide scale spallation mainly due to the preferential internal oxidation of Ti and Nb at the grain boundaries (supplementary **Fig. S16-S18**.).

In conclusion, we have successfully inherited the natural mixing characteristics amongst refractory elements in achieving RHEAs with promising tensile properties even at the as-cast state. We reveal that the interesting mechanical behavior of the naturally-predicted $Ti_{38}V_{15}Nb_{23}Hf_{24}$ RHEA is due to a rare β' precipitation strengthening effect, and a corresponding dislocation channeling mechanism that causes the destruction of these nano-scale precipitates. By varying the overall composition in this alloy system, e.g., Ti/V ratio, matrix phase stability and precipitate size can be altered, leading to different property combinations. More importantly, this study demonstrates that having such a nature-provided alloy design hint, drastically expedites the discovery of alloys that exhibit superior mechanical and physicochemical properties.

**Acknowledgements**

The TEM analyses were accomplished at the Materials Research Science and Engineering Center (MRSEC) shared experimental facilities at Massachusetts Institute of Technology, financially supported by the National Science Foundation (NSF) under the grant No. DMR-1419809. The synchrotron X-ray diffraction experiments were carried out on beamline 11ID-C at the Argon National Laboratory, Chicago, USA. S.J.K. and E.S.P. acknowledge the financial support from the Creative Materials Discovery Program through the National Research Foundation of Korea (NRF) funded by the Ministry of Science and ICT (No. NRF-2019M3D1A1079213) and the Institute of Engineering Research at Seoul National University. T.F. acknowledges the financial support from JSPS KAKENHI under the Grant No. JP18H05456 (Grant-in-Aid for Scientific Research on Innovative Areas 2018-2023). Y.J.Z. and T.F. thanks Mr. Kunio Shinbo for technical support on APT measurement and Prof. Goro Miyamoto for valuable discussions. S.L.W. and C.C.T. express their sincere gratitude to Profs. Donald R. Sadoway, Ju Li; Drs. Daniel B. Miracle, Feng He, Yang Ren, Hyunseok Oh, Jinwoo Kim, Shao-Shi Rui; and Mr. Minseok Kim for their consistent support and inputs.

**Author contributions**

S.L.W. and C.C.T. conceptualized the project and designed the research; S.L.W. was the leading research scientist of this work; S.L.W., S.J.K., and E.S.P. fabricated the RHEA ingots; J.Y.K. performed the synchrotron X-ray diffraction experiments; S.L.W. and Y.Z. carried out the TEM characterizations; Y.J.Z. and T.F. performed the APT measurements; S.L.W. and C.C.T. analyzed the data and wrote the paper; All authors discussed the results and approved the final version of the manuscript.

**Methods**

*Alloy fabrication*

The nine-component equiatomic TiVCrZrNbMoHfTaW master RHEA for composition search was fabricated by arc-melting from pure elements under an Ar atmosphere (purity higher than 99.9 wt. %). Guided by the quantitative elemental partitioning analyses *via* EDS (details provided in the following section), the designed $Ti_{38}V_{15}Nb_{23}Hf_{24}$ at. % RHEA and its equiatomic counterpart ($Ti_{25}V_{25}Nb_{25}Hf_{25}$ at. %) were both produced through arc-melting followed by suction casting. During the arc-melting process, all the RHEA ingots were flipped upside down and re-melted at least five times with a piece of Ti getter being employed to prevent potential oxygen contamination. The two latter RHEAs were subsequently wrapped by Ta foils and sealed in quartz tubes filled with sponge Ti before being homogenized at 1100 °C for 10 h. The as-homogenized specimens were further cold-rolled down to 30 % thickness reduction, recrystallized at 1000 °C for 5 min under Ar protection, and then water-quenched to ambient temperature.

*Microstructural and structural characterization*

Meso-scale microstructural characterizations including morphological observation, electron channeling contrast imaging, EBSD, and EDS analyses were all performed in a TESCAN MIRA 3 SEM. EBSD data were quantitatively post-processed in an orientation imaging microscopy (OIM) software. Specimens for EBSD and EDS investigations were sectioned from bulk RHEA ingots *via* electrical discharge machining (EDM) followed by mechanical grinding and polishing to achieve mirror-finish surface conditions. Synchrotron X-ray diffraction experiments for crystal structure and phase constitution assessments were conducted on beamline 11ID-C at the Argon National Laboratory, Chicago, USA. Two-dimensional diffractograms were acquired by utilizing a high-resolution X-ray radiation source with 0.1173 Å wavelength and were post-analyzed in a GASA-II software. An FEI Tecnai G2 Spirit TWIN TEM and a JEOL JEM-2010F HRTEM were employed for micro- and nano-scale fine structure characterizations. Thin-foil TEM specimens were prepared from RHEA pieces before and after deformation by mechanically ground down to ~20 μm thickness before being ion-milled. Atomic-scale elemental distribution was investigated using a CAMECA LEAP 4000 HR APT under voltage pulsing mode (detection efficiency 37 %). The APT specimens were prepared by an FEI Quanta 3D focused ion beam milling system. All the APT measurements were carried out at a specimen temperature of 80 K with a pulse fraction and rate of 20 % and 200 kHz, while the raw data were post-analyzed in an IVAS ver. 3.6 software.

*Mechanical property measurement*

Rectangular dog-bone-shaped tensile specimens with gauge geometry of 6.5×2.5×1 mm$^3$ were sectioned from bulk RHEA specimens using EDM. Before being tested, the specimens were mechanically ground and polished for post-mortem surface steps observation, and speckle patterns were coated on their surfaces for DIC analyses (frame rate 2 Hz). Ambient-temperature uniaxial tensile testing was conducted on a Gatan micro-mechanical testing platform at a strain rate of 1×10$^{-3}$ s$^{-1}$. At least three specimens were tested for each microstructural condition to ensure reproducibility. Local strain evolution profiles during tensile experiments were calculated in a GOM software (https://www.gom.com/3d-software/gom-correlate.html) to determine the incipience of necking.

**References**

1. Cantor, B., Chang, I. T. H., Knight, P. & Vincent, A. J. B. Microstructural development in equiatomic multicomponent alloys. *Mater. Sci. Eng. A* **375–377**, 213–218 (2004).
2. Yeh, J. W. *et al.* Nanostructured High-Entropy Alloys with Multiple Principal Elements: Novel Alloy Design Concepts and Outcomes. *Adv. Eng. Mater.* **6**, 299–303 (2004).
3. Li, Z., Pradeep, K. G., Deng, Y., Raabe, D. & Tasan, C. C. Metastable high-entropy dual-phase alloys overcome the strength-ductility trade-off. *Nature* **534**, 227–230 (2016).
4. Sydney, U. *et al.* A Fracture-Resistant High-Entropy Alloy for Cryogenic Applications. *Science (80-. ).* **345**, 1153–1159 (2014).
5. George, E. P., Raabe, D. & Ritchie, R. O. High entropy alloys. *Nat. Rev. Mater.* **4**, 515–534 (2019).

6. Yang, T. *et al.* Multicomponent intermetallic nanoparticles and superb mechanical behaviors of complex alloys. *Science (80-. ).* (2018). doi:10.1126/science.aas8815

7. Jo, Y. H. *et al.* Cryogenic strength improvement by utilizing room-temperature deformation twinning in a partially recrystallized VCrMnFeCoNi high-entropy alloy. *Nat. Commun.* (2017). doi:10.1038/ncomms15719

8. Ding, J., Yu, Q., Asta, M. & Ritchie, R. O. Tunable stacking fault energies by tailoring local chemical order in CrCoNi medium-entropy alloys. *Proc. Natl. Acad. Sci. U. S. A.* (2018). doi:10.1073/pnas.1808660115

9. Varvenne, C., Leyson, G. P. M., Ghazisaeidi, M. & Curtin, W. A. Solute strengthening in random alloys. *Acta Mater.* **124**, 660–683 (2017).

10. Li, Q.-J., Sheng, H. & Ma, E. Strengthening in multi-principal element alloys with local-chemical-order roughened dislocation pathways. *Nat. Commun.* (2019). doi:10.1038/s41467-019-11464-7

11. Oh, H. S. *et al.* Engineering atomic-level complexity in high-entropy and complex concentrated alloys. *Nat. Commun.* **10**, 1–8 (2019).

12. Senkov, O. N., Miracle, D. B., Chaput, K. J. & Couzinie, J. P. Development and exploration of refractory high entropy alloys - A review. *Journal of Materials Research* (2018). doi:10.1557/jmr.2018.153

13. Fu, X., Schuh, C. A. & Olivetti, E. A. Materials selection considerations for high entropy alloys. *Scr. Mater.* (2017). doi:10.1016/j.scriptamat.2017.03.014

14. Senkov, O. N., Wilks, G. B., Miracle, D. B., Chuang, C. P. & Liaw, P. K. Refractory high-entropy alloys. *Intermetallics* (2010). doi:10.1016/j.intermet.2010.05.014

15. Miracle, D. B. & Senkov, O. N. A critical review of high entropy alloys and related concepts. *Acta Materialia* (2017). doi:10.1016/j.actamat.2016.08.081

16. Qi, L. & Chrzan, D. C. Tuning ideal tensile strengths and intrinsic ductility of bcc refractory alloys. *Phys. Rev. Lett.* (2014). doi:10.1103/PhysRevLett.112.115503

17. Peterson N.L. Diffusion in refractory metals. (1960).

18. Distefano, J. R., Pint, B. A. & Devan, J. H. Oxidation of refractory metals in air and low pressure oxygen gas. *Int. J. Refract. Met. Hard Mater.* (2000). doi:10.1016/S0263-4368(00)00026-3

19. Lei, Z. *et al.* Enhanced strength and ductility in a high-entropy alloy via ordered oxygen complexes. *Nature* **563**, 546–550 (2018).

20. Huang, H. *et al.* Phase-Transformation Ductilization of Brittle High-Entropy Alloys via Metastability Engineering. *Adv. Mater.* **29**, 1–7 (2017).

21. Wang, Y., Li, J., Hamza, A. V & Barbee, T. W. Ductile crystalline – amorphous nanolaminates. *Proc. Natl. Acad. Sci.* **104**, 11155–11160 (2007).

22. Hutchinson, J. W. & Neale, K. W. Influence of strain-rate sensitivity on necking under uniaxial tension. *Acta Metall.* **25**, 839–846 (1977).

23. Kato, H., Ozu, T., Hashimoto, S. & Miura, S. Cyclic stress-strain response of superelastic Cu-Al-Mn alloy single crystals. *Mater. Sci. Eng. A* (1999).

24. Lilensten, L. *et al.* On the heterogeneous nature of deformation in a strain-transformable beta metastable Ti-V-Cr-Al alloy. *Acta Mater.* (2019). doi:10.1016/j.actamat.2018.10.003

25. Gu, X. F., Furuhara, T. & Zhang, W. Z. PTCLab: Free and open-source software for calculating phase transformation crystallography. *J. Appl. Crystallogr.* (2016). doi:10.1107/S1600576716006075

26. V́ıtek, V. Intrinsic stacking faults in body-centred cubic crystals. *Philos. Mag.* **18**, 773–786 (1968).

27. Olfe, J. & Neuhäuser, H. Dislocation groups, multipoles, and friction stresses in α‐CuZn alloys. *Phys. status solidi* (1988). doi:10.1002/pssa.2211090115

28. Lai, M. J., Tasan, C. C. & Raabe, D. Deformation mechanism of ω-enriched Ti-Nb-based gum metal: Dislocation channeling and deformation induced ω-β transformation. *Acta Mater.* **100**, 290–300 (2015).

29. Chen, W. *et al.* Origin of the ductile-to-brittle transition of metastable β-titanium alloys: Self-hardening of ω-precipitates. *Acta Mater.* (2019). doi:10.1016/j.actamat.2019.03.034

# Natural-mixing guided design of refractory high-entropy alloys with as-cast tensile ductility

Shaolou Wei[1], Sang Jun Kim[2], Ji Yun Kang[1], Yong Zhang[3], Yongjie Zhang[4],

Tadashi Furuhara[4], Eun Soo Park[2], and Cemal Cem Tasan[1,*]

[1.] *Department of Materials Science and Engineering, Massachusetts Institute of Technology, Cambridge, MA 02139, USA*
[2.] *Department of Materials Science and Engineering, Seoul National University, Seoul 08826, Republic of Korea*
[3.] *Materials Research Laboratory, Massachusetts Institute of Technology, Cambridge, MA 02139, USA*
[4.] *Institute for Materials Research, Tohoku University, Sendai 980-8577, Japan*

*Corresponding authors: Cemal Cem Tasan (tasan@mit.edu)

## Contents

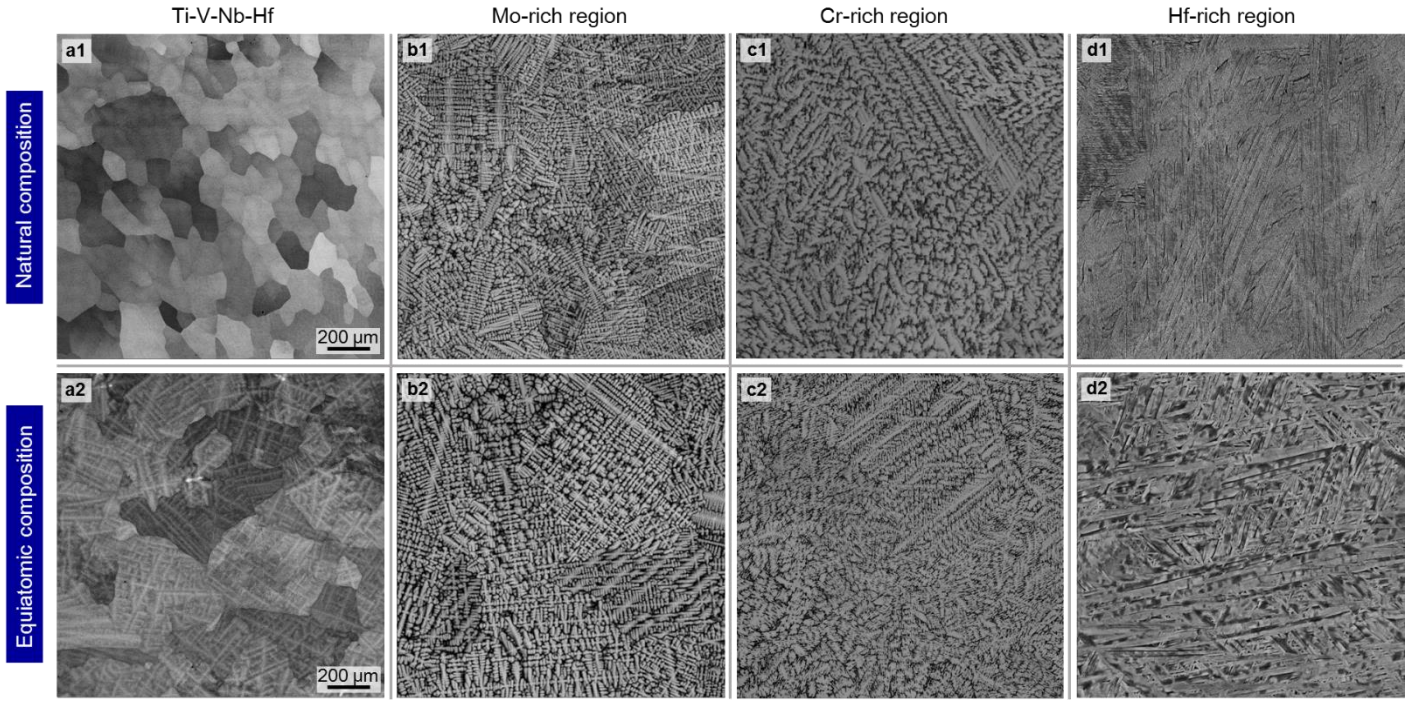

**Fig. S1| As-cast microstructures of natural-mixing-suggested alloys and their equiatomic variants. a1-d1,** three RHEAs and one refractory medium-entropy alloy (RMEA) designed from the composition measured from the four predominant phase-separated regions (main text **Fig. 1 a, b**). **a2-d2,** equiatomic compositional variants with respect to **a1-d1**. It is revealed that the $Ti_{38}V_{15}Nb_{23}Hf_{24}$ composition inherited from the largest single-phase region in the nine-component master REHA (main text **Fig. 1 a**) not only exhibits the minimum amount of casting features but also demonstrates a near-equiaxed grain morphology.

**Tab. S1| EDS analyses of the four predominant phase-separated regions in the nine-component master RHEA.** As addressed in the main text, in the follow-up compositional design process, we consider a compositional threshold of 10.00 at. % in determining principal alloying elements. In order to minimize the interference from potential systematic drifting, all the EDS measurements were taken from a $1.0 \times 1.0$ μm$^2$ square region in each individual area. In the following table, principal elements and their portions applied in achieving the alloys presented in **Fig. S1 a1-d1** are highlighted in blue. We also note here that the minor presence of other alloying elements in each region can also be considered as the hint for compositional optimization for accelerating future RHEAs design.

| Alloying element (at. %) | Ti-V-Nb-Hf region | Mo-rich region | Cr-rich region | Hf-rich region |
|---|---|---|---|---|
| Ti | 29.50 | | | |
| V | 12.04 | | | |
| Cr | 4.54 | | | |
| Zr | 5.29 | | | |
| Nb | 17.65 | *N.A. at the moment* | *N.A. at the moment* | *N.A. at the moment* |
| Mo | 4.25 | | | |
| Hf | 18.77 | | | |
| Ta | 5.39 | | | |
| W | 2.57 | | | |
| As-cast microstructure | **Fig. S1 a1** | **Fig. S1 b1** | **Fig. S1 c1** | **Fig. S1 d1** (RMEA) |

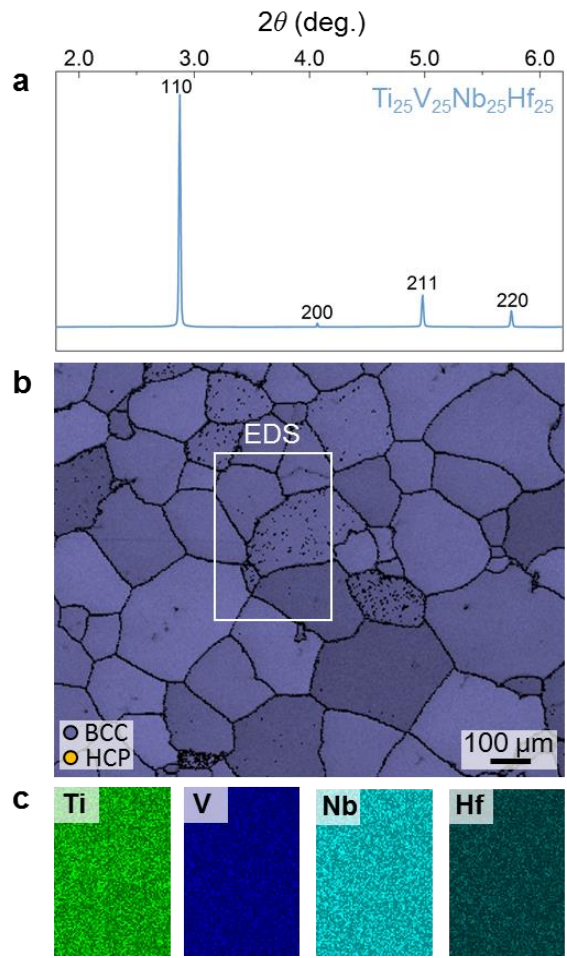

**Fig. S2| Meso-scale microstructural characterization of the equiatomic $Ti_{25}V_{25}Nb_{25}Hf_{25}$ RHEA. a,** synchrotron X-ray diffraction patterns indicate that such an equiatomic variant also demonstrates a BCC structure with a lattice parameter $a =$ 3.299 Å. **b,** EBSD phase map confirming the single-phase constitution; **c,** EDS elemental mapping taken across multiple grain boundaries proves that after homogenization and recrystallization processing, the four principal alloying elements demonstrate a spatially uniform distribution. Similar to the meso-scale characterization results for the $Ti_{38}V_{15}Nb_{23}Hf_{24}$ at. % RHEA (main text **Fig. 1 c**), at the resolution limits of EBSD and synchrotron X-ray, no clear trait of either elemental inhomogeneity or presence of secondary phases are detected.

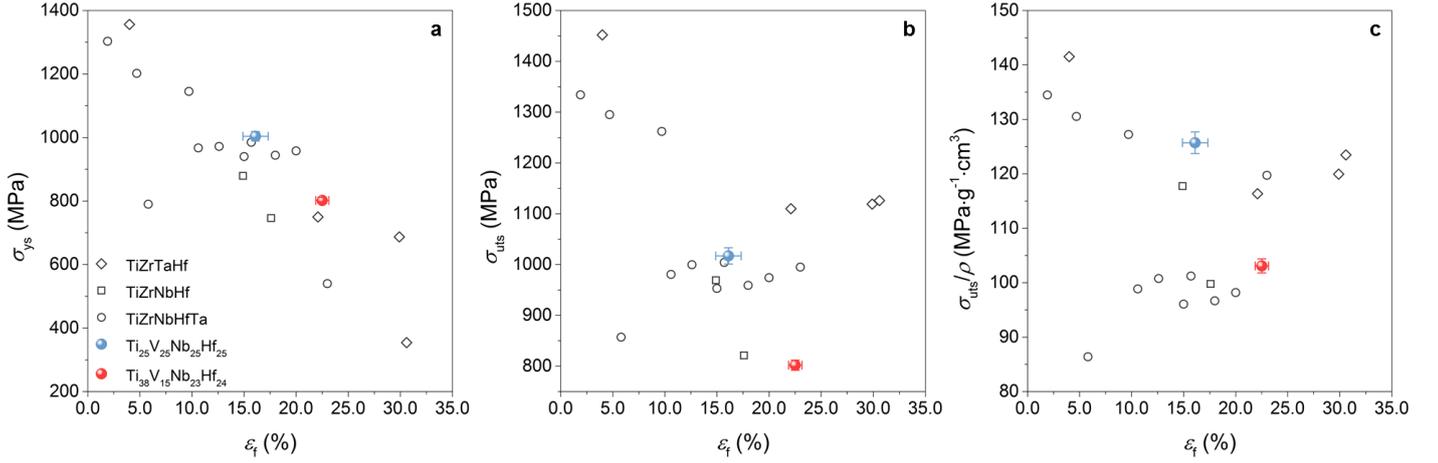

**Fig. S3| Complementary Ashby charts for strength-ductility comparisons.** To enable comprehensive assessments of the mechanical performances for the designed $Ti_{38}V_{15}Nb_{23}Hf_{24}$ at its equiatomic variant, we provide here three additional Ashby charts regarding: **a,** yield strength ($\sigma_{ys}$) versus fracture elongation ($\varepsilon_f$); **b,** ultimate tensile strength ($\sigma_{uts}$) versus fracture elongation; and **c,** specific ultimate tensile strength ($\sigma_{uts}/\rho$) versus fracture elongation. As mentioned in the main text, in these groups of comparisons we only consider reported RHEAs consisting purely of refractory elements. We also recognize that more recent study has revealed the introduction of non-transition metallic elements (Al, as an example[1–3]) or unique interstitial dopants[4] into RHEAs can also bring about intriguing tensile performances. Due to page limit in the main text, we summarize in the following **Tab. S2** all the detailed referencing information for the above comparisons.

**Tab. S2| Tensile properties of RHEAs reported in the literature.** $\rho$, $\dot{\varepsilon}$, $\sigma_y$, $\sigma_{uts}$, and $\varepsilon_f$ represent density, strain rate of mechanical testing, yield strength, ultimate tensile strength, and fracture elongation for the surveyed RHEAs. Superscripts (i) and (ii) denote HCP-phase formed through thermal quenching and mechanically-induced martensitic transformation.

| Alloy system | Composition | $\rho$, g/$cm^3$ | Phase | $\dot{\varepsilon}$, $s^{-1}$ | $\sigma_y$, MPa | $\sigma_{uts}$, MPa | $\varepsilon_f$, % | Ref. |
|---|---|---|---|---|---|---|---|---|
| TiZrTaHf | TiZrTaHf | 10.26 | BCC | $1\times10^{-3}$ | 1356±86 | 1452±88 | 4.0±1.7 | 5 |
| | $TiZrTa_{0.6}Hf$ | 9.54 | BCC+HCP (i) | $1\times10^{-3}$ | 750±80 | 1110±85 | 22.1±1.6 | 5 |
| | $TiZrTa_{0.5}Hf$ | 9.33 | BCC+HCP (i) | $1\times10^{-3}$ | 687±23 | 1119±39 | 29.9±2.7 | 5 |
| | $TiZrTa_{0.4}Hf$ | 9.12 | BCC+HCP (i), (ii) | $1\times10^{-3}$ | 354±26 | 1126±55 | 30.6±1.7 | 5 |
| TiZrNbHf | TiZrNbHf | 8.23 | BCC | $1\times10^{-3}$ | 879 | 969 | 14.9 | 6 |
| | TiZrNbHf | 8.23 | BCC | $2\times10^{-4}$ | 746 | 821 | 17.6 | 4 |
| TiZrNbHfTa | TiZrNbHfTa | 9.92 | BCC | $5\times10^{-3}$ | 790 | 857 | 5.8 | 7 |
| | TiZrNbHfTa | 9.92 | BCC | $1\times10^{-3}$ | 1202 | 1295 | 4.7 | 8 |
| | TiZrNbHfTa | 9.92 | BCC | $1\times10^{-3}$ | 1303 | 1334 | 1.9 | 8 |
| | TiZrNbHfTa | 9.92 | BCC | $1\times10^{-3}$ | 1145 | 1262 | 9.7 | 8 |
| | TiZrNbHfTa | 9.92 | BCC | $1\times10^{-3}$ | 958 | 974 | 20.0 | 9 |
| | TiZrNbHfTa | 9.92 | BCC | $1\times10^{-3}$ | 944 | 959 | 18.0 | 9 |
| | TiZrNbHfTa | 9.92 | BCC | $1\times10^{-3}$ | 940 | 953 | 15.0 | 9 |
| | TiZrNbHfTa | 9.92 | BCC | $2\times10^{-4}$ | 985 | 1004 | 15.7 | 10 |
| | TiZrNbHfTa | 9.92 | BCC | $2\times10^{-4}$ | 972 | 999.5 | 12.6 | 10 |
| | TiZrNbHfTa | 9.92 | BCC | $2\times10^{-4}$ | 967 | 980.6 | 10.6 | 10 |
| | $Ti_{1.3}ZrNb_{0.2}HfTa_{0.2}$ | 8.31 | BCC+HCP (ii) | $1\times10^{-4}$ | 540 | 995 | 23.0 | 11 |

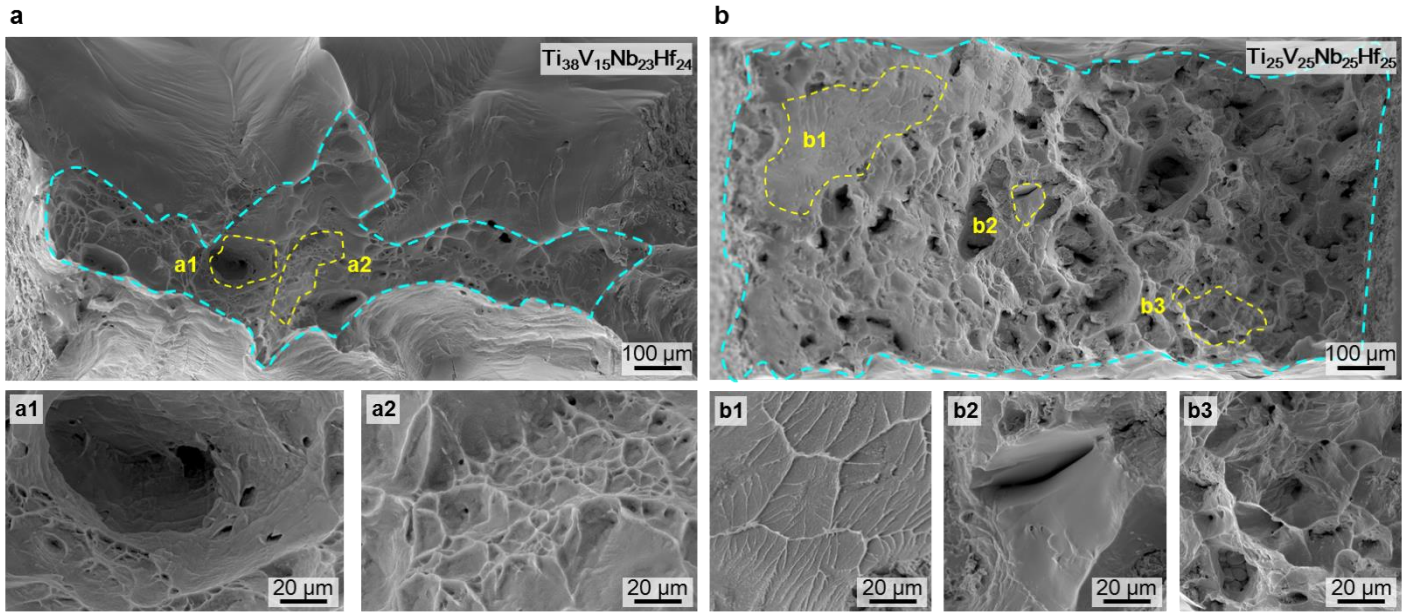

**Fig. S4| Fractography of the Ti$_{38}$V$_{15}$Nb$_{23}$Hf$_{24}$ RHEA and its equiatomic variant after tensile testing. a** and **b,** overviews of the fracture surfaces at lower magnifications indicate that although ductile fracture occurs in both RHEAs, the comparatively more significant fracture surface area reduction (see the dashed cyan lines in **a** and **b** as a guide) also suggests the more pronounced post-necking elongation of the Ti$_{38}$V$_{15}$Nb$_{23}$Hf$_{24}$ RHEA. **a1** and **a2,** magnified micrographs of representative areas in **a**. **b1-b3,** in addition to typical features for ductile fracture, a small portion of cleavage fracture is also observed in the Ti$_{25}$V$_{25}$Nb$_{25}$Hf$_{25}$ RHEA which can also exhibit a minor contribution to its less eminent post-necking elongation.

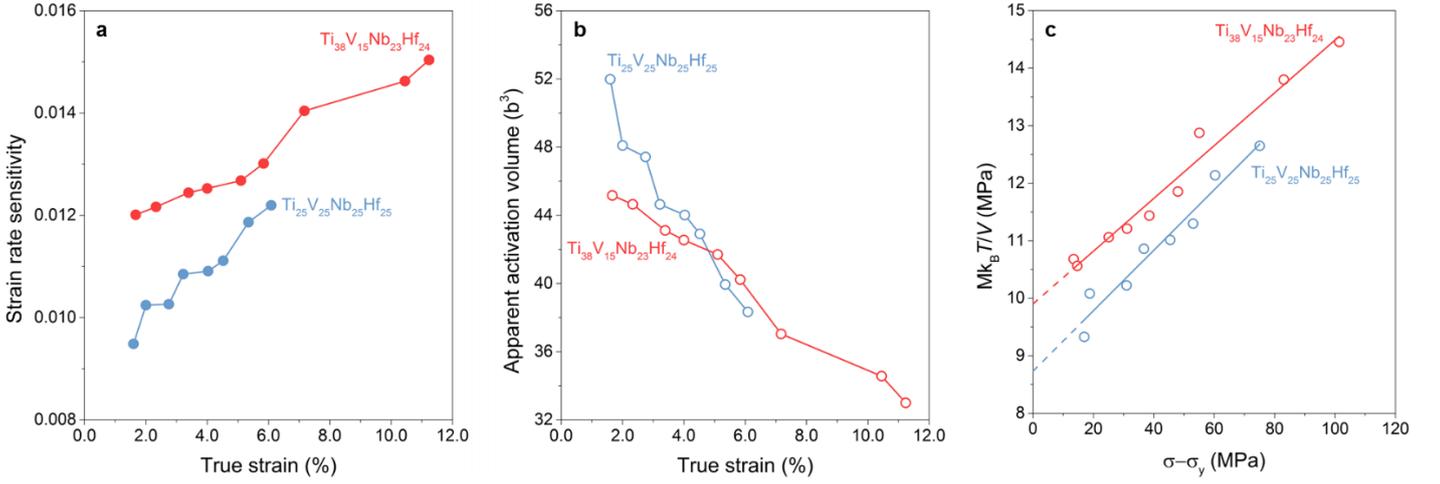

**Fig. S5| Assessments of thermal activation parameters for stable plastic flow.** According to the non-linear instability analyses by Hutchinson and Neale[12], strain rate sensitivity[13,14] ($m = \partial \ln\sigma/\partial \ln\dot{\varepsilon}|_T$) can exhibit a pronounced influence on post necking elongation of metallic alloys at a certain temperature $T$. To elucidate the post-necking elongation difference between $Ti_{38}V_{15}Nb_{23}Hf_{24}$ and $Ti_{25}V_{25}Nb_{25}Hf_{25}$ RHEAs and to also shed a bit more light on the underlying deformation mechanisms, we have performed strain rate jump tests (alternating jumps between $1\times10^{-4}$ and $1\times10^{-3}$ $s^{-1}$) to reveal the thermal activation parameters at ambient temperature. **a,** throughout the uniform elongation regions, it is clear that the $Ti_{38}V_{15}Nb_{23}Hf_{24}$ RHEA exhibits a much larger strain rate sensitivity than its equiatomic variant. Such a characteristic indicates that after the onset of localized deformation (necking), the $Ti_{38}V_{15}Nb_{23}Hf_{24}$ RHEA will exhibit a more evident hardenability, rendering a comparatively more eminent post-necking elongation. **b,** measured apparent activation volume ($V_{app.}$) with respect to increasing true plastic strain. By definition[15], the apparent activation volume $V_{app.} = Mk_BT(\partial \ln\dot{\varepsilon}/\partial \sigma)|_{T,\mu_s}$ signifies both the mobile dislocation density and the stress-dependent dislocation density at a certain temperature $T$ and microstructural state $\mu_s$. In the above equation, $M$ represents the Taylor factor which relates the uniaxial stress and the resolved shear stress for polycrystalline metals demonstrating an equiaxed grain morphology. In our $V_{app.}$ calculations, an $M$ value of 2.75 was adopted (pencil glide with no preferred slip system[16]), which has been widely accepted for BCC-structure metals[17,18], dilute[19] and concentrated alloys [20,21]. In line with the literature that reported activation volumes for concentrated alloys with FCC[22] and BCC[20,21] structures, the measured $V_{app.}$ values in our study also decrease as a function of increasing plastic strain, which is mostly ascribed to the decreasing dislocation spacing ($l = \sqrt{\rho_{disloc.}}$) as plastic straining proceeds. Since both these alloys were tested at fully recrystallized states, the $V_{app.}$ values at the very initial states of plastic flow indicate the interactions between solutes or any other potential heterogeneity within the microstructure. A comparatively larger $V_{app.}$ value ~52 $b^3$ can be seen in the $Ti_{25}V_{25}Nb_{25}Hf_{25}$ RHEA, implying the potential heterogeneous sites (later proved to be nano-precipitates) exhibit larger dimensions than the $Ti_{38}V_{15}Nb_{23}Hf_{24}$ RHEA (see **Fig. S11** and **Fig. 4 b** in the main text). **c,** a more quantitative analysis, the Haasen plot[13], which reveals the dependency between the deviated flow stress $\sigma - \sigma_{YS}$ and the quantity $Mk_BT/V_{app.} = (\partial\sigma/\partial\ln\dot{\varepsilon})|_{T,\mu_s}$ is also presented. The larger positive offset at the y-axis in the $Ti_{38}V_{15}Nb_{23}Hf_{24}$ RHEA implies the presence of a stronger integrated effect of solute and precipitation strengthening (although separating these two contribution requires accurate prediction of solid solution strength in BCC-structured RHEAs). The slope of the two RHEAs almost perverse at a comparable level, validating their similar forest hardening behavior (also see **Fig. 2 b** in the main text).

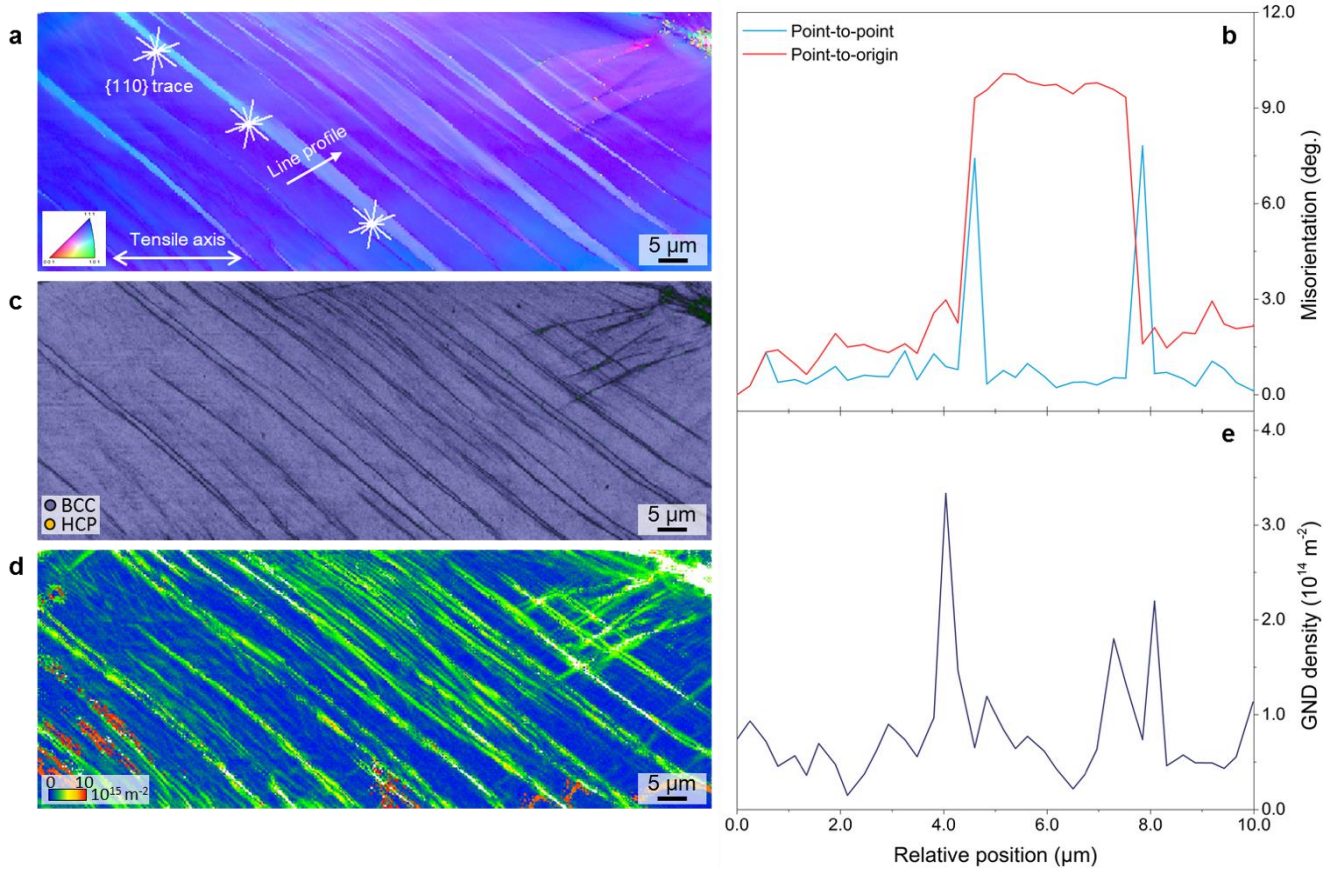

**Fig. S6| Validation of meso-scale deformation mechanisms of the $Ti_{38}V_{15}Nb_{23}Hf_{24}$ RHEA at a local strain level of ~40.0 %. a,** IPF map indicates the presence of multiple rectilinear bands with moderate orientation deviation with respect to their adjacent matrix. **b,** a quantitative line profile (position highlighted in **a**) confirms that unlike the characteristic misorientation difference generated by either mechanically-induced martensitic transformation[5,11,23] or deformation twinning[24,25] in BCC-structured alloys, there only exists a maximum ~10 deg. deviation across the band. Coinciding with the deformation banding mechanisms revealed in β-Ti alloys, a much more moderate misorientation angle ~3 deg. is also observed within the channeling band or its adjacent matrix. These quantitative analyses signify that the $Ti_{38}V_{15}Nb_{23}Hf_{24}$ RHEA presumably maintain mechanically stable even at an extensive local strain level for which the dislocation-mediated plasticity is the dominant deformation module. **c,** overlapped IQ and phase map validates the monitored region exhibit a single-phase BCC structure although the interface between channeling bands and matrix reveals a decreased IQ value. **d,** geometrically necessary dislocation (GND) density map calculated by adopting the algorithm developed by Pantleon[26]. **e,** GND density line profile confirms the dislocation-mediated plasticity event is highly confined within individual channeling bands (GND develops more significantly at the interface).

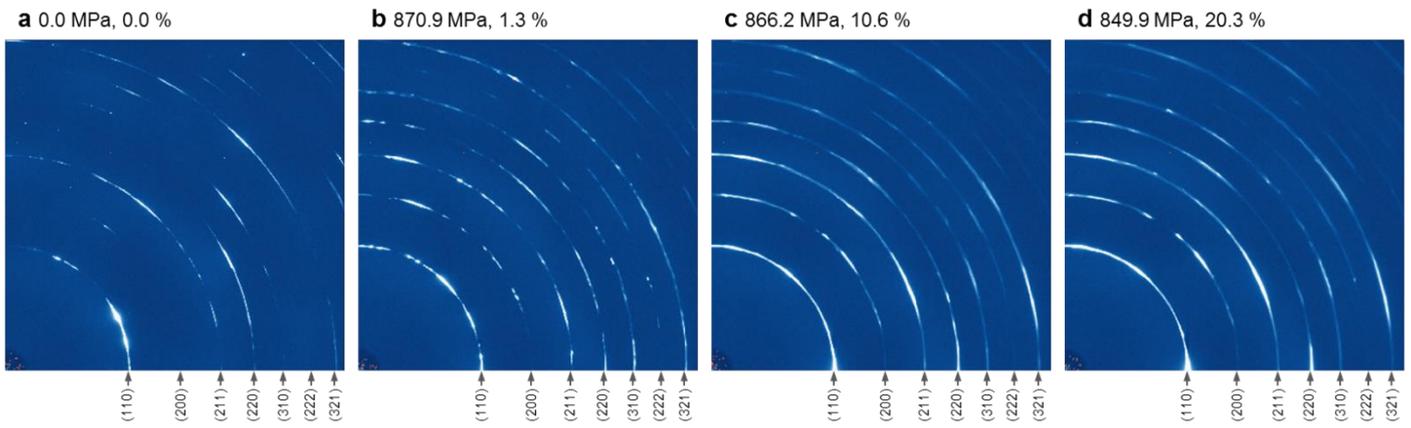

**Fig. S7| In-situ tensile testing under synchrotron X-ray for mechanical stability validation of the $Ti_{38}V_{15}Nb_{23}Hf_{24}$ RHEA.** In order to accurately examine the mechanical stability of the major BCC-phase in the $Ti_{38}V_{15}Nb_{23}Hf_{24}$ RHEA and to also complement the EBSD measurements, we have performed in-situ tensile tests with synchrotron X-ray diffraction (beamline 11 ID-C, Argonne National Laboratory, $\lambda = 0.1173$ Å, $\dot{\varepsilon} = 1 \times 10^{-3}\ s^{-1}$, guage geometry 8.0×2.0×1.0 mm³). **a-d,** Two-dimensional diffractograms taken as a function of increasing deformation level confirm that there is no trait of mechanically-assisted α' or α" (orthorhombic)-martensitic transformation[24,27,28] until macroscopic failure takes place.

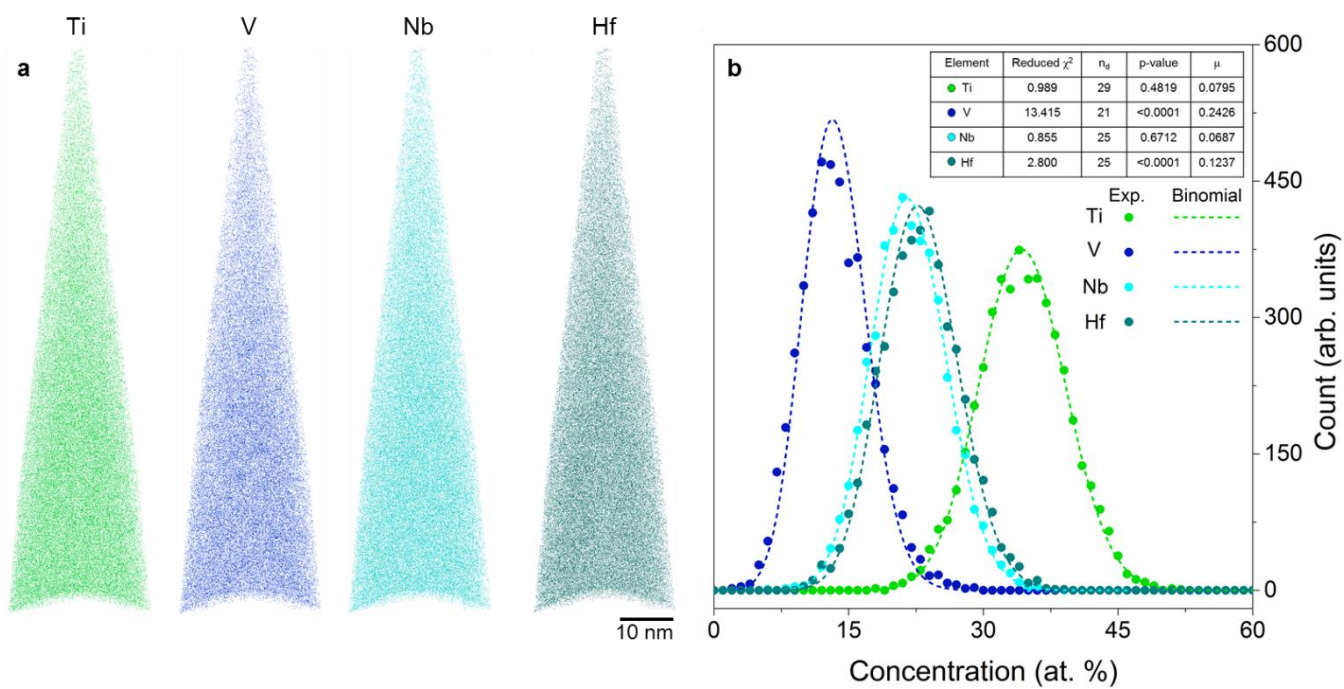

**Fig. S8| APT mapping and binomial analysis for the homogenized-recrystallized $Ti_{38}V_{15}Nb_{23}Hf_{24}$ RHEA. a,** raw elemental mapping results for Ti, V, Nb, and Hf. **b,** the raw data in **a** were subjected to quantitative binomial regression analysis[29], validating the presence of a stronger segregation (clustering) trend in V compared to the other elements. On the basis of **b**, we construct the 20 at. % V isocomposition surface (main text **Fig. 4 b** and **c**) to reveal chemical heterogeneity brought about by nano-scale BCT-structured precipitates (main text **Fig. 4 a**).

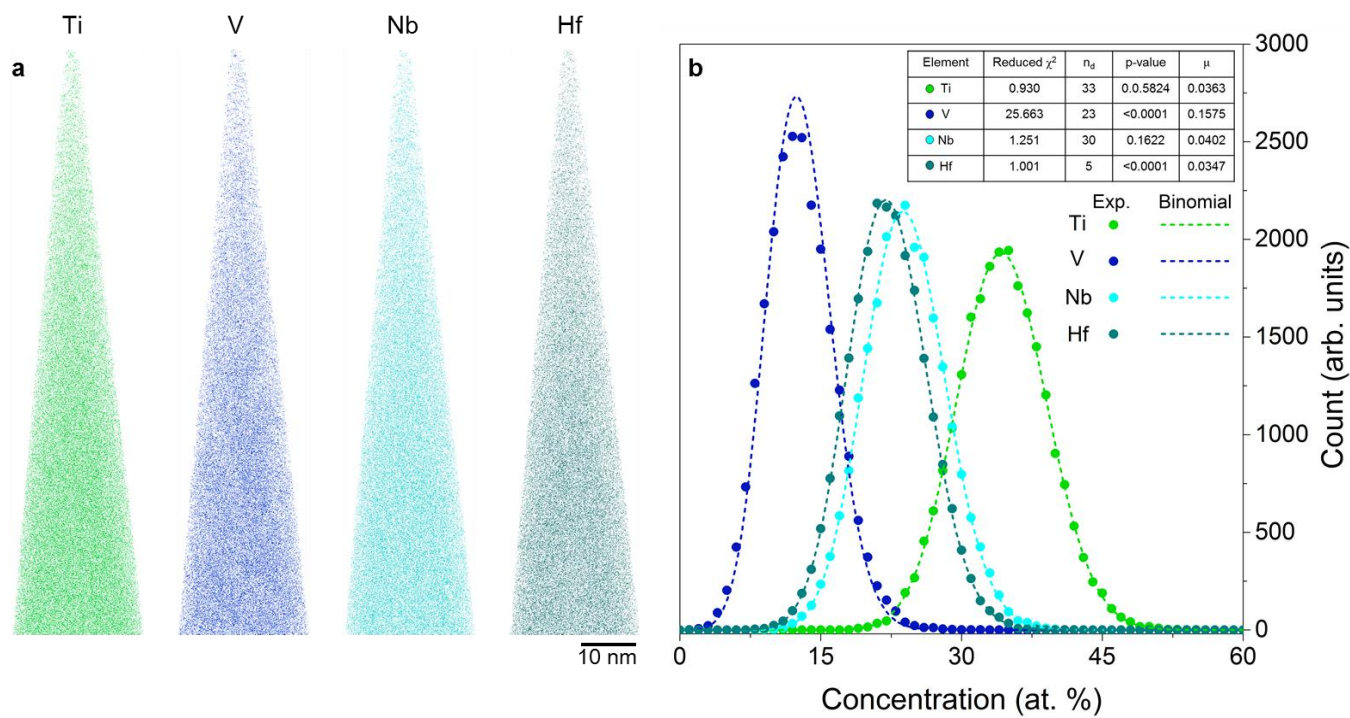

**Fig. S9| APT mapping and binomial analysis for the as-cast Ti$_{38}$V$_{15}$Nb$_{23}$Hf$_{24}$ RHEA.** As shown in **Fig. S1** and **Fig. 2** in the main text, the Ti$_{38}$V$_{15}$Nb$_{23}$Hf$_{24}$ RHEA not only demonstrates the minimum amount of casting segregation amongst all the examined RHEAs and MEA but also preserves an almost invariant mechanical response after extensive thermomechanical treatment (see "**Methods**" in the main text). By elucidating the underlying deformation mechanism across multiple length-scales, we argue that the observed nano-sized precipitates (main text **Fig. 4**) is presumably an intrinsic compositional characteristic and exhibit decent thermal stability. **a,** raw elemental mapping results for Ti, V, Nb, and Hf. **b,** to validate the postulate we have performed the binomial regression analysis of which the results highlight the similar V segregation (clustering) effect.

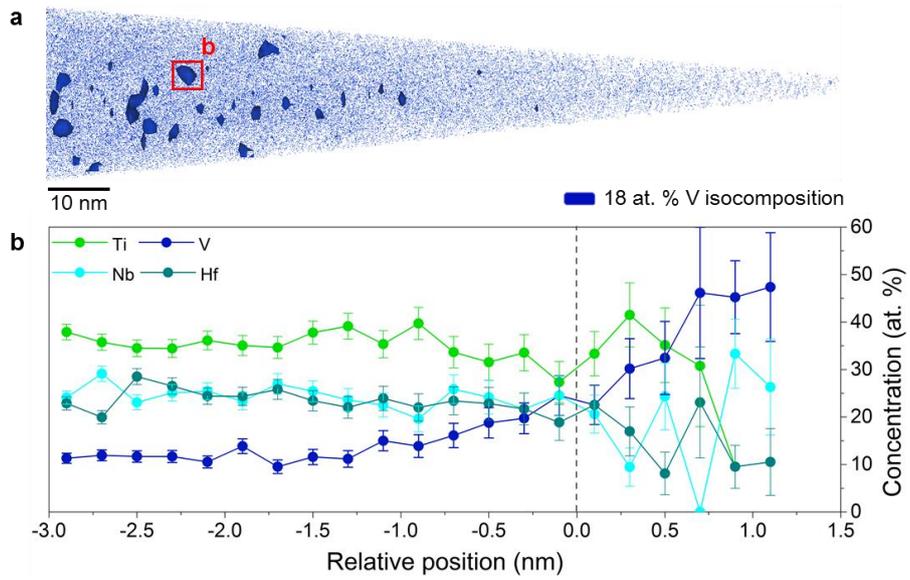

**Fig. S10| Confirmation of nano-precipitates within the as-cast $Ti_{38}V_{15}Nb_{23}Hf_{24}$ RHEA. a,** based on the binomial analysis discussed in **Fig. S9** an 18 at. % V isocomposition surface is constructed to validate the existence of nano-precipitates; **b,** the corresponding proxigram taken at a selected site in **a** shows the elemental partitioning features between precipitate and its adjacency.

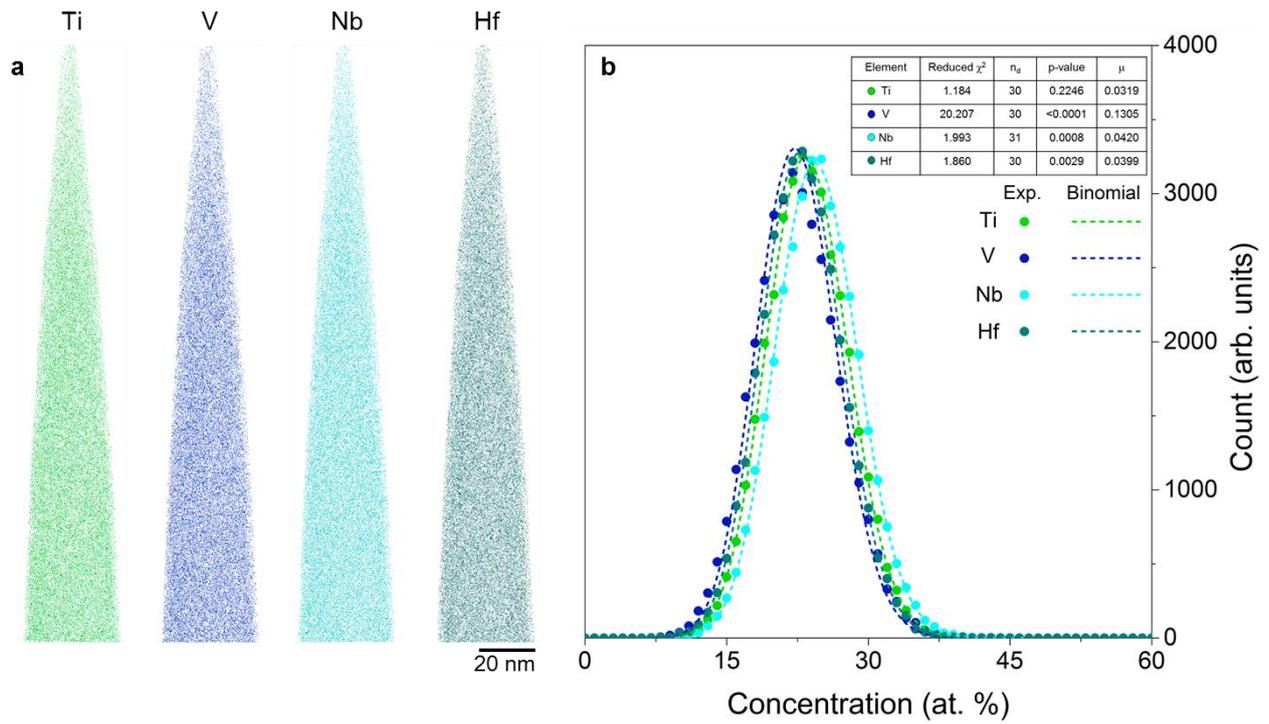

**Fig. S11| APT mapping and binomial analysis for the homogenized-recrystallized $Ti_{25}V_{25}Nb_{25}Hf_{25}$ RHEA.** To complement the discussion addressed in the main text regarding the distinctive yield strength and yielding inflection observed in the equiatomic variant, we provide here the APT measurement to validate the existence of nano-scale precipitates in larger dimensions and higher volumetric content. **a,** raw elemental mapping results for Ti, V, Nb, and Hf. **b,** the corresponding binomial regression analysis results.

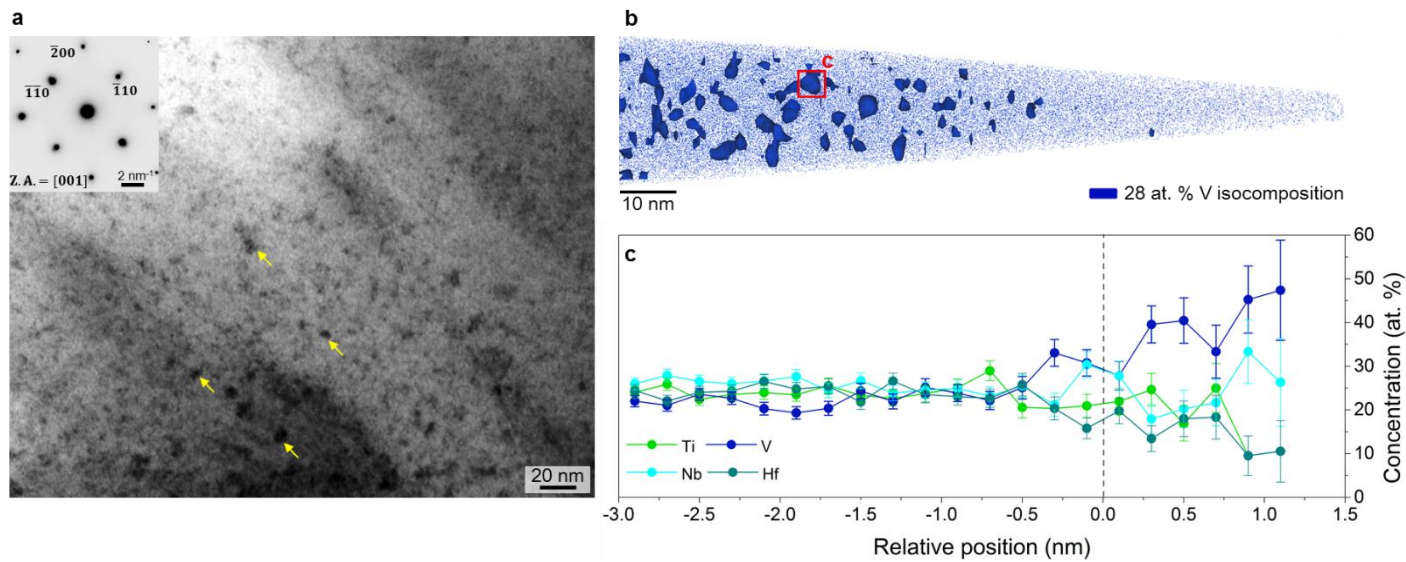

**Fig. S12| Confirmation of nano-precipitates within the homogenized-recrystallized $Ti_{25}V_{25}Nb_{25}Hf_{25}$ RHEA. a,** TEM-BF micrograph taken with respect to [001] zone axis clearly reveals the presence of precipitates, although their size and content are not sufficiently high to be resolved by the corresponding SAED patterns (inset of **a**). **b,** based on the analysis in **Fig. S11 b** a 28 at. % V isocomposition surface is provided which validates the nano-precipitates in this equiatomic variant appear in larger dimensions and higher contents. **c,** proxigram acquired at a selected site in **b** demonstrates the elemental partitioning features between precipitate and its adjacency.

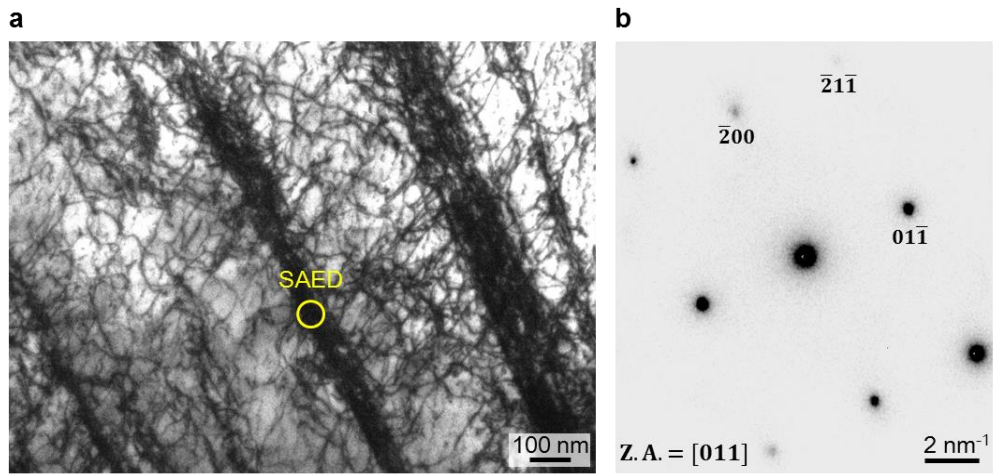

**Fig. S13| TEM micrographs of a highly deformed $Ti_{38}V_{15}Nb_{23}Hf_{24}$ RHEA. a,** BF image corresponds to **Fig. 4 e** in the main text; **b,** SAED pattern taken with respect to the $[011]$ zone axis proves the absence of either martensitic phases or twins within the highly dislocated region.

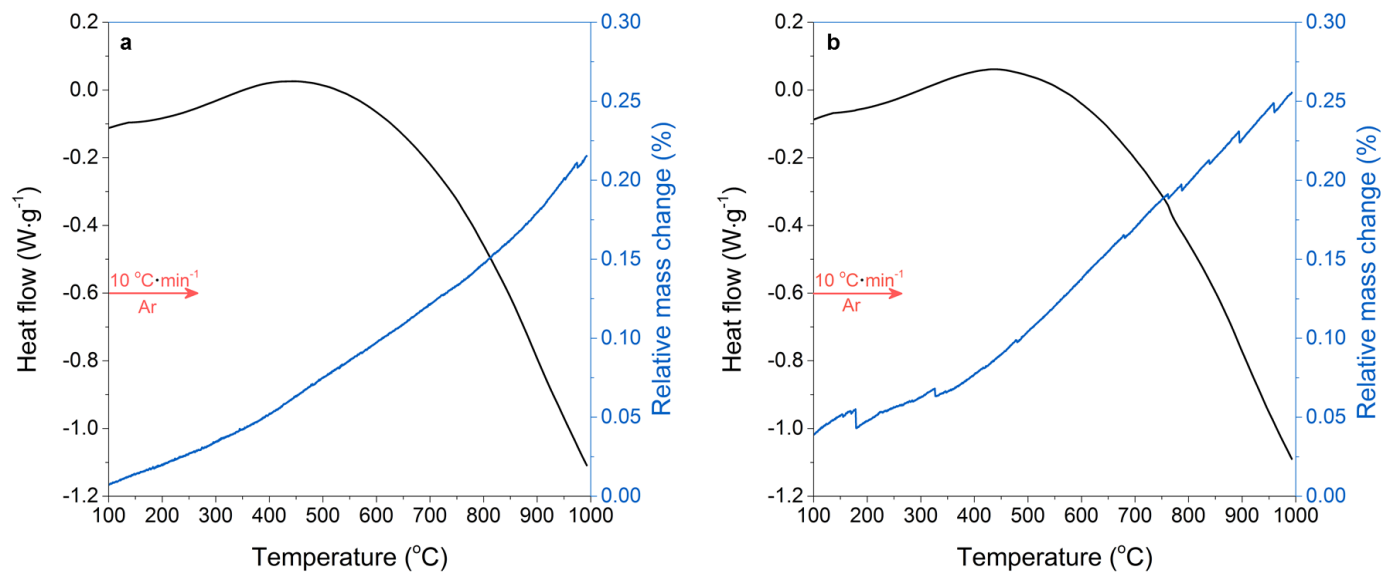

**Fig. S14| TG-DSC analyses of phase stability for the two RHEAs. a,** $Ti_{38}V_{15}Nb_{23}Hf_{24}$ RHEA. **b,** $Ti_{25}V_{25}Nb_{25}Hf_{25}$ RHEA. The coupled thermal analyses were performed on a METLLER TOLEDO TG-DSC facility under the protection of inert Ar flow. As seen in **a** and **b**, in the testing temperature range from 100-1000 °C no trait of any evident endothermic or exothermic peak is detected, indicating no eminent phase transformation has taken place.

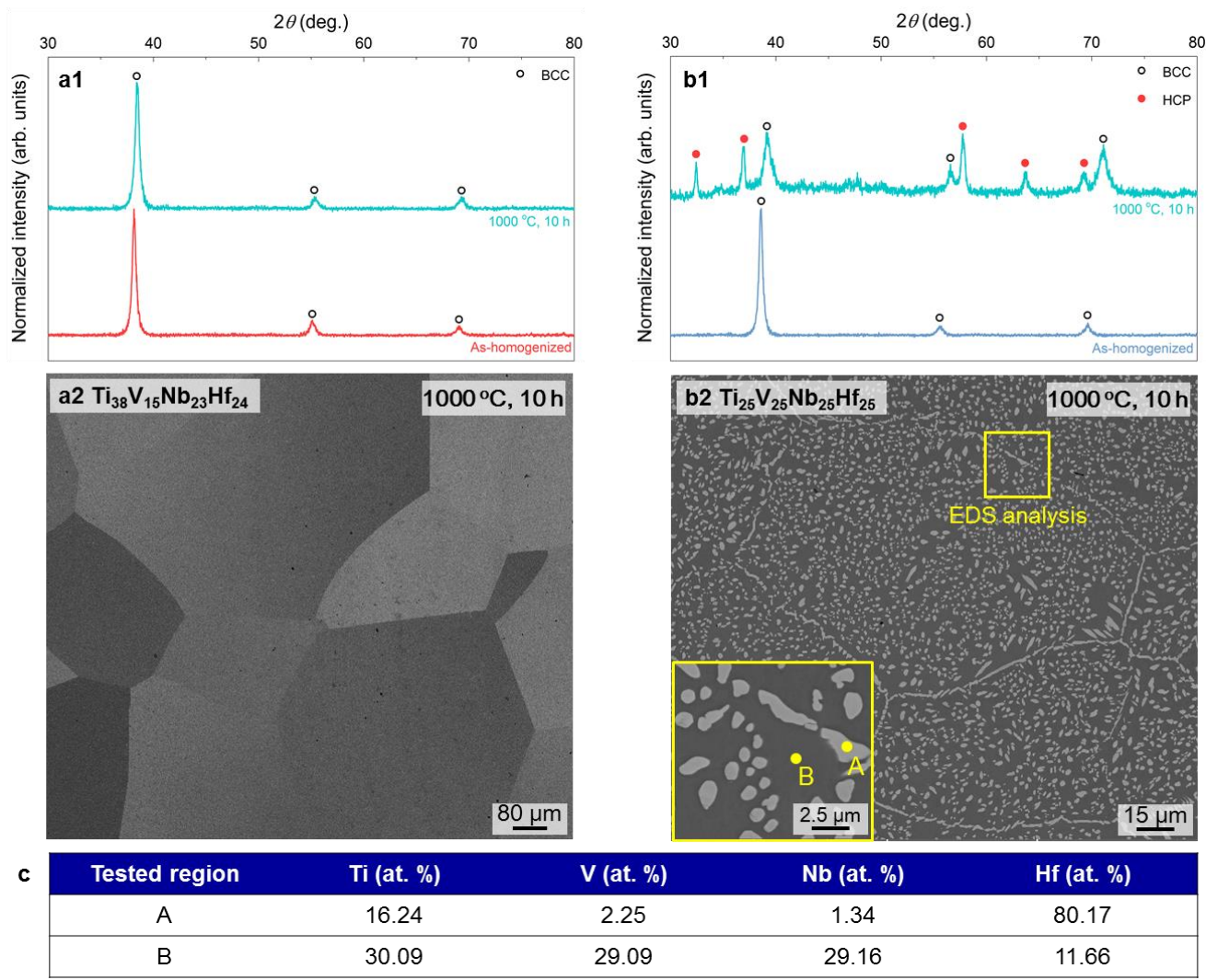

**Fig. S15| Phase stability of the two REHAs assessed at 1000 ºC.** In order to shed a bit more light on the compositional stability for the $Ti_{38}V_{15}Nb_{23}Hf_{24}$ RHEA and its equiatomic variant, we have examined their phase stability at 1000 ºC by conducting a 10 h annealing treatment under protective conditions (see "**Methods**" in the main text). Phase constitutions for the annealed specimens were assessed by a Bruker D2 PHASWER X-ray diffractometer with a monochromatic Cu-Kα radiation source. **a1** and **b1,** after being annealed at 1000 ºC for 10 h, the $Ti_{38}V_{15}Nb_{23}Hf_{24}$ RHEA still preserves a single-BCC microstructure at the meso-scale, while in contrast, diffraction peaks of an HCP-phase are clearly resolved in its equiatomic counterpart. **a2** and **b2,** BSE micrographs indicate that the $Ti_{38}V_{15}Nb_{23}Hf_{24}$ RHEA has only gone through grain growth after annealing, whereas a significant amount of precipitates has nucleated and coarsened within the $Ti_{25}V_{25}Nb_{25}Hf_{25}$ RHEA (even forming networks along the grain boundaries). The inset of **b2** reveals a magnified BSE micrograph for the selected area being examined by EDS point analyses. **c,** EDS results indicate that the HCP-structured precipitation evidently enriches in Ti and Hf but depleted in V and Nb compared with its BCC-structure matrix. The above analyses validate that the $Ti_{38}V_{15}Nb_{23}Hf_{24}$ composition inherited from the natural mixing characteristics amongst refractory elements does exhibit a more desirable compositional stability than its equiatomic variant.

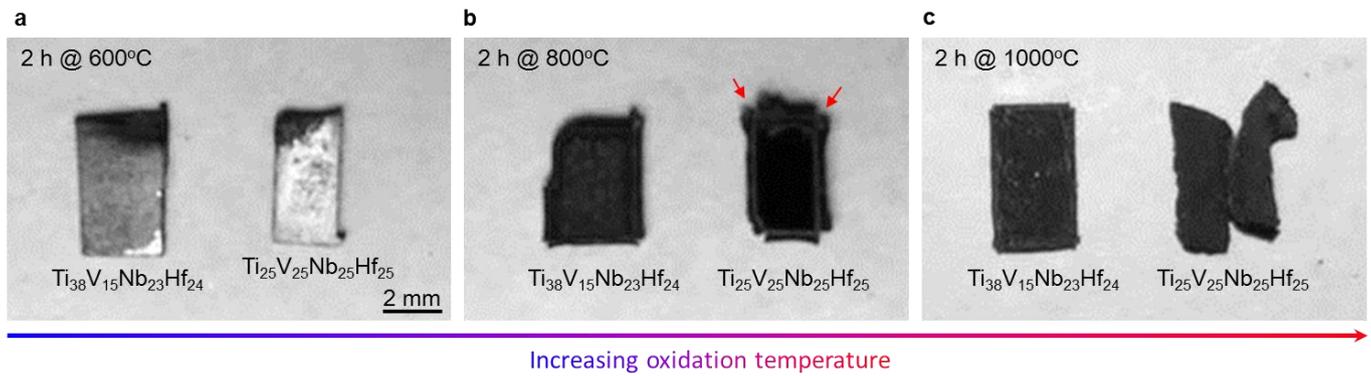

**Fig. S16| Assessments of oxidation resistance at 600, 800, and 1000 °C for 2 h.** Although complete understandings of oxidation resistance do require long-term kinetic analyses and sometimes involve more aggressive conditions [30], the initial stage of thermal oxidation can still be employed to rationally elucidate the underlying mechanisms [31], especially when early-stage catastrophic oxidation occurs. After being oxidized at 600, 800, and 1000 °C for 2 h it is recognized from the macroscopic morphologies (comparatively shown in **a**, **b**, and **c**) that the $Ti_{38}V_{15}Nb_{23}Hf_{24}$ RHEA demonstrates decent adherence between oxide scale and its substrate with almost no mechanical spallation taking place even at 1000 °C. In contrast, evident spallation is seen in the $Ti_{25}V_{25}Nb_{25}Hf_{25}$ RHEA at 800 °C (see the red arrows). When oxidation temperature rises up to 1000 °C, serious catastrophic oxidation takes place and completely destructs the specimen. To shed a bit more light on the origin of the distinctive difference in oxidation resistance, we have examined the cross section of the two specimens oxidized at 800 °C (**Figs. S17** and **S18**).

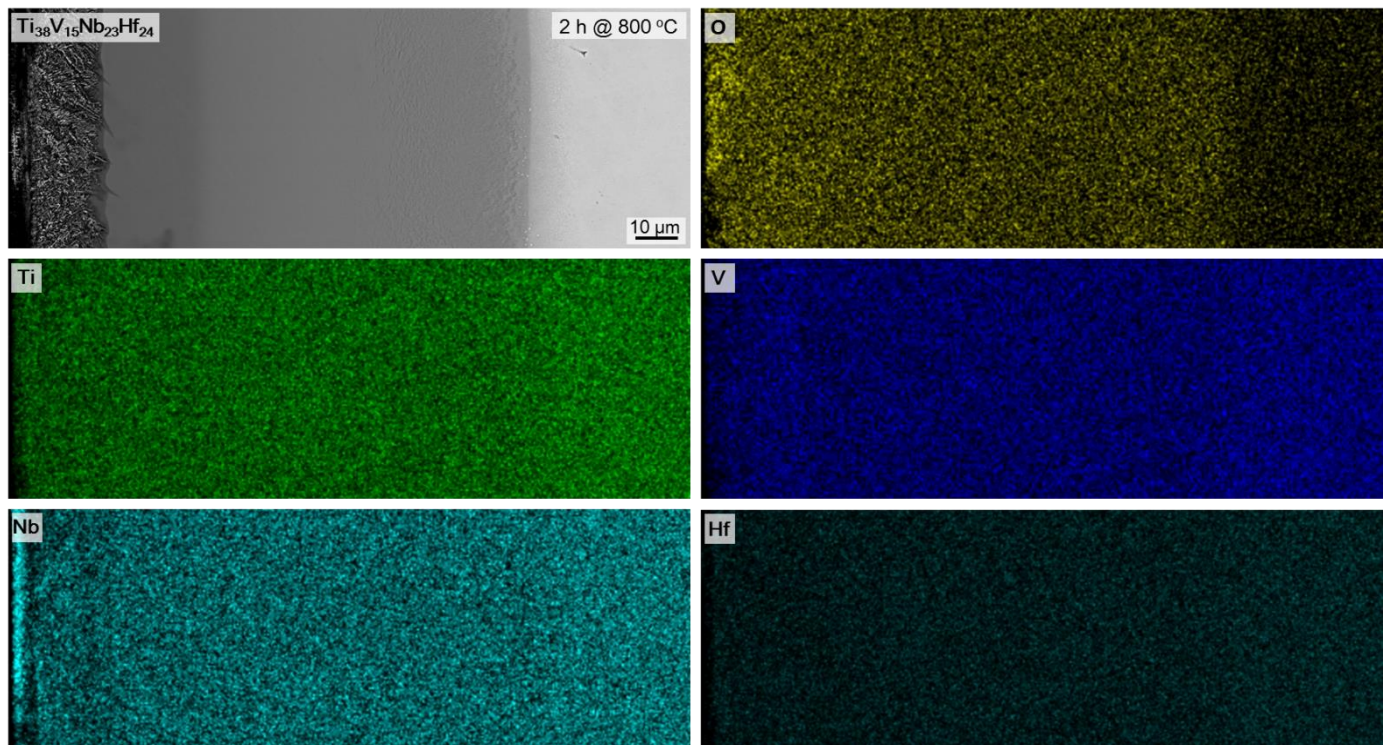

**Fig. S17| Cross-sectional analyses of the oxidized Ti$_{38}$V$_{15}$Nb$_{23}$Hf$_{24}$ RHEA.** As seen from the BSE micrograph, the Ti$_{38}$V$_{15}$Nb$_{23}$Hf$_{24}$ RHEA develops a relatively dense oxide scale with ~15 μm thickness together with two layers of inter-diffusion regions after being subjected to 800 °C 2 h oxidation. No trait of micro-cracking is observed at the interface between the oxide scale and the substrate that lies beneath. EDS elemental mappings highlight that Nb enriches more significantly at the outermost portion of the oxide scale, and no site-preferred elemental partitioning (such as preferential oxidation at grain boundaries) is observed.

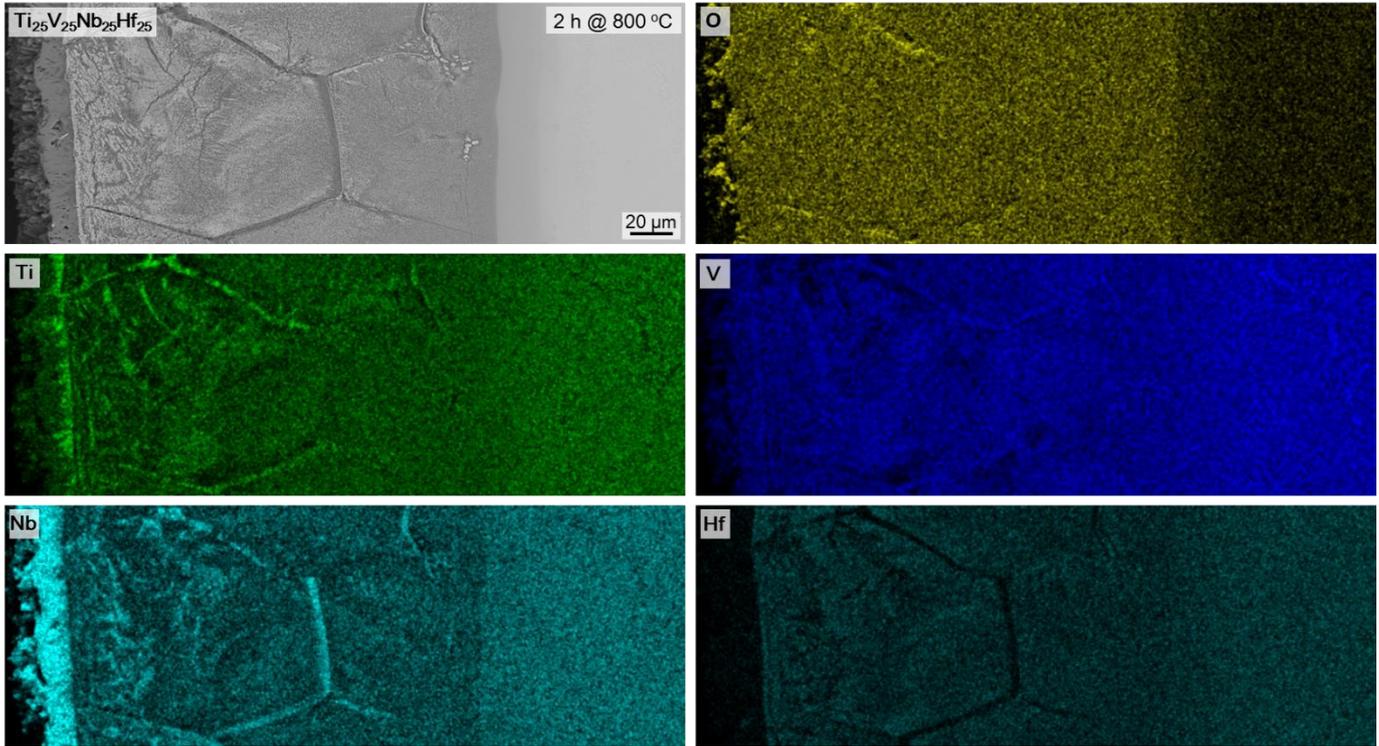

**Fig. S18| Cross-sectional analyses of the oxidized $Ti_{25}V_{25}Nb_{25}Hf_{25}$ RHEA.** In contrast to the decent protective oxide scale observed in the $Ti_{38}V_{15}Nb_{23}Hf_{24}$ RHEA under the same oxidation condition, a far more complicated microstructure exists in the oxidized $Ti_{25}V_{25}Nb_{25}Hf_{25}$ RHEA. The most striking feature is the presence of significant preferential oxidation of Nb and Ti, especially along the grain boundaries. This kind of oxidation mechanism will inevitably result in drastic growth stress due to the volumetric differences between oxides and the matrix alloy (characterized by the Pilling-Bedworth ratio [32,33], $PBR$). In particular, the $PBR$ value of Nb at its most thermodynamically stable state reaches 2.69, indicating a strong propensity for chipping off, which as a result, facilitating macroscopic mechanical spallation confirmed in **Fig. S16 b** and **c**. It is also recognized that oxygen penetration depth is much larger than that seen in the $Ti_{38}V_{15}Nb_{23}Hf_{24}$ RHEA, implying the developed oxide scale is unlikely to suppress internal oxidation. In addition, a significant amount of Nb enriches at the outmost oxide with a very thin layer enriching in Ti developed underneath. Comparisons amongst **Figs. S16**, **S17**, and **S18** indicate that the catastrophic oxidation in the equiatomic $Ti_{25}V_{25}Nb_{25}Hf_{25}$ RHEA is mainly attributed to the destructive preferential oxidation of Nb and Ti along the grain boundaries together with the more significant oxygen penetration which expediate internal oxidation.

**Supplementary references**

1. Wang, L. *et al.* Ductile Ti-rich high-entropy alloy controlled by stress induced martensitic transformation and mechanical twinning. *Mater. Sci. Eng. A* (2019). doi:10.1016/j.msea.2019.138147
2. Wu, Y. *et al.* Phase stability and mechanical properties of AlHfNbTiZr high-entropy alloys. *Mater. Sci. Eng. A* **724**, 249–259 (2018).
3. Wang, L. *et al.* Superelastic effect in Ti-rich high entropy alloys via stress-induced martensitic transformation. *Scr. Mater.* (2019). doi:10.1016/j.scriptamat.2018.10.035
4. Lei, Z. *et al.* Enhanced strength and ductility in a high-entropy alloy via ordered oxygen complexes. *Nature* **563**, 546–550 (2018).
5. Huang, H. *et al.* Phase-Transformation Ductilization of Brittle High-Entropy Alloys via Metastability Engineering. *Adv. Mater.* **29**, 1–7 (2017).
6. Wu, Y. D. *et al.* A refractory Hf 25 Nb 25 Ti 25 Zr 25 high-entropy alloy with excellent structural stability and tensile properties. **130**, 277–280 (2014).
7. Dirras, G. *et al.* Elastic and plastic properties of as-cast equimolar TiHfZrTaNb high-entropy alloy. *Mater. Sci. Eng. A* **654**, 30–38 (2016).
8. Senkov, O. N. & Semiatin, S. L. Microstructure and properties of a refractory high-entropy alloy after cold working. *J. Alloys Compd.* **649**, 1110–1123 (2015).
9. Juan, C. C. *et al.* Simultaneously increasing the strength and ductility of a refractory high-entropy alloy via grain refining. *Mater. Lett.* **184**, 200–203 (2016).
10. Chen, S. *et al.* Grain growth and Hall-Petch relationship in a refractory HfNbTaZrTi high-entropy alloy. *J. Alloys Compd.* **795**, 19–26 (2019).
11. Lilensten, L. *et al.* Design and tensile properties of a bcc Ti-rich high-entropy alloy with transformation-induced plasticity. *Mater. Res. Lett.* **5**, 110–116 (2017).
12. Hutchinson, J. W. & Neale, K. W. Influence of strain-rate sensitivity on necking under uniaxial tension. *Acta Metall.* **25**, 839–846 (1977).
13. Haasen, P. Plastic deformation of nickel single crystals at low temperatures. *Philos. Mag.* **3**, 384–418 (1958).
14. Mecking, H. & Kocks, U. F. Kinetics of flow and strain-hardening. *Acta Metall.* (1981). doi:10.1016/0001-6160(81)90112-7
15. Argon, A. *Strengthening Mechanisms in Crystal Plasticity*. *Strengthening Mechanisms in Crystal Plasticity* **9780198516**, (2007).
16. Kocks, U. F. The relation between polycrystal deformation and single-crystal deformation. *Metall. Mater. Trans.* (1970). doi:10.1007/BF02900224
17. Kapoor, R. & Nemat-Nasser, S. I. A. Comparison between high and low strain-rate deformation of tantalum. *Metall. Mater. Trans. A Phys. Metall. Mater. Sci.* (2000).
18. Hosseini, E. & Kazeminezhad, M. Dislocation structure and strength evolution of heavily deformed tantalum. *Int. J. Refract. Met. Hard Mater.* (2009). doi:10.1016/j.ijrmhm.2008.09.006
19. Lassila, D. H., Goldberg, A. & Becker, R. The effect of grain boundaries on the athermal stress of tantalum and tantalum-tungsten alloys. *Metall. Mater. Trans. A Phys. Metall. Mater. Sci.* (2002). doi:10.1007/s11661-002-0333-9
20. Couziné, J. P. *et al.* On the room temperature deformation mechanisms of a TiZrHfNbTa refractory high-entropy alloy. *Mater. Sci. Eng. A* **645**, 255–263 (2015).
21. Lilensten, L. *et al.* Study of a bcc multi-principal element alloy: Tensile and simple shear properties and underlying deformation mechanisms. *Acta Mater.* (2018). doi:10.1016/j.actamat.2017.09.062
22. Laplanche, G., Bonneville, J., Varvenne, C., Curtin, W. A. & George, E. P. Thermal activation parameters of plastic flow reveal deformation mechanisms in the CrMnFeCoNi high-entropy alloy. *Acta Mater.* **143**, 257–264 (2018).
23. Lee, B. S. *et al.* Stress-induced α″ martensitic transformation mechanism in deformation twinning of metastable β-type Ti-27Nb-0.5Ge alloy under tension. *Mater. Trans.* (2016). doi:10.2320/matertrans.M2016208
24. Castany, P., Yang, Y., Bertrand, E. & Gloriant, T. Reversion of a Parent 310 α″ Martensitic Twinning System at the Origin of {332} 113 ⓒβ Twins Observed in Metastable β Titanium Alloys. *Phys. Rev. Lett.* **117**, 1–6 (2016).
25. Liu, H. *et al.* Abnormal Deformation Behavior of Oxygen-Modified β-Type Ti-29Nb-13Ta-4.6Zr Alloys for

Biomedical Applications. *Metall. Mater. Trans. A Phys. Metall. Mater. Sci.* (2017). doi:10.1007/s11661-016-3836-5

26. Pantleon, W. Resolving the geometrically necessary dislocation content by conventional electron backscattering diffraction. *Scr. Mater.* **58**, 994–997 (2008).
27. Mantani, Y., Takemoto, Y., Hida, M., Sakakibara, A. & Tajima, M. Phase transformation of α″ martensite structure by aging in Ti-8mass%Mo alloy. *Mater. Trans.* (2004). doi:10.2320/matertrans.45.1629
28. Chang, L. L., Wang, Y. D. & Ren, Y. In-situ investigation of stress-induced martensitic transformation in Ti-Nb binary alloys with low Young's modulus. *Mater. Sci. Eng. A* (2016). doi:10.1016/j.msea.2015.11.005
29. Zhang, Y., Miyamoto, G. & Furuhara, T. Atom Probe Compositional Analysis of Interphase Precipitated Nano-Sized Alloy Carbide in Multiple Microalloyed Low-Carbon Steels. *Microsc. Microanal.* (2019). doi:10.1017/S1431927618015374
30. Wei, S. L., Huang, L. J., Li, X. T., An, Q. & Geng, L. Interactive effects of cyclic oxidation and structural evolution for Ti-6Al-4V/(TiC+TiB) alloy composites at elevated temperatures. *J. Alloys Compd.* **752**, 164–178 (2018).
31. Wei, S. *et al.* Sub-stoichiometry-facilitated oxidation kinetics in a δ-TixC-doped Ti-based alloy. *npj Mater. Degrad.* **3**, 3 (2019).
32. Bedworth, R. E. & Pilling, N. B. The oxidation of metals at high temperatures. *J Inst Met* (1923).
33. Xu, C. & Gao, W. Pilling-bedworth ratio for oxidation of alloys. *Mater. Res. Innov.* (2000). doi:10.1007/s100190050008